\documentclass[a4paper,11pt]{article}

\usepackage{PackageBispectrum}

\begin{document}
\def\thefootnote{\fnsymbol{footnote}}

\begin{center}
\Large{\textbf{Cosmic shear bispectrum \\from second-order perturbations in General Relativity}}
\\[0.5cm]
\large{Francis Bernardeau$^a$, Camille Bonvin$^{b,c}$,\\ Nicolas Van de Rijt$^a$ and Filippo Vernizzi$^a$}
\\[0.5cm]

\small{
\textit{{}$^a$CEA, Institut de Physique Th{\'e}orique, 91191 Gif-sur-Yvette c\'edex, France\\ CNRS, URA-2306, 91191 Gif-sur-Yvette c\'edex, France}\\
\vspace{0.2cm}
\textit{$^b$Kavli Institute for Cosmology Cambridge and Institute of Astronomy, \\
Madingley Road, Cambridge CB3 OHA, UK}\\
\vspace{0.2cm}
 \textit{$^c$DAMTP, Centre for Mathematical Sciences, Wilberforce Road, Cambridge CB3 OWA, UK}
}
\end{center}

\vspace{1cm}

\hrule \vspace{0.3cm}
\noindent \small{\textbf{Abstract}} \\[0.3cm]
\noindent
Future lensing surveys will be nearly full-sky and reach an unprecedented depth, probing scales closer and closer to the Hubble radius. This motivates the study of the cosmic shear beyond the small-angle approximation and including general relativistic corrections that are usually suppressed on sub-Hubble scales. The complete expression of the reduced cosmic shear at second order including all relativistic effects was derived in~\cite{Bernardeau:2009bm}. 
In the present paper we compute the resulting cosmic shear bispectrum when all these effects are properly taken into account and we compare it to primordial non-Gaussianity of the local type. 
The new general relativistic effects are generically smaller than the standard non-linear couplings. However, their relative importance increases at small multipoles and for small redshifts of the sources. The dominant effect among these non standard corrections is due to the inhomogeneity of the source redshift. In the squeezed limit, its amplitude can become of the order of the standard couplings when the redshift of the sources is below $0.5$.
Moreover, while  the standard non-linear couplings depend on the angle between the short and long mode, the relativistic corrections do not and overlap almost totally with local type non-Gaussianity. We find that they can contaminate the search for a primordial local signal by $f_{\rm NL}^{\rm loc} \gtrsim 10$.
\vspace{0.5cm} \hrule
\def\thefootnote{\arabic{footnote}}
\setcounter{footnote}{0}

\parskip 0pt
\vspace{1cm}

\vspace{1cm}


\section{Introduction}

With the advent of future wide-field surveys, weak gravitational lensing will become a premier probe of cosmology and an important tool to constrain dark energy, neutrinos and the initial conditions (see for instance \cite{2010AnP...522..324D,Giannantonio:2011ya}). In order to fully exploit the potentiality of the  convergence and shear fields, it will be important to use their whole statistics. In particular, the lensing bispectrum represents a complementary probe to the power spectrum, as it will provide constraints  that are comparable to those obtained from the power spectrum alone \cite{Berge:2009xj}.

One of the primary interests of wide-field surveys is to look for primordial non-Gaussianities generated in the early universe and the lensing bispectrum represents a natural tool to capture such features \cite{Bernardeau:2002fc,Takada:2003sv,2011MNRAS.411..595P,Jeong:2011rh,2011arXiv1107.1656S}. Given the current and forecasted constraints on $\fNL$, there is a plethora of second-order effects intervening between the initial conditions and the observations that could be potentially relevant. Some of these effects, such as the second-order growth of matter fluctuations in Eulerian perturbation theory, the Born correction, the lens-lens coupling and the non-linear conversion between the galaxy shape distortion and the (observable) reduced shear, have been thoroughly studied in the past years and are expected to dominate  the lensing 3-point statistics on small-angular scales \cite{1997A&A...322....1B,Cooray:2000uu,2002A&A...389L..28B,Cooray:2002mj,2003MNRAS.344..857T,Dodelson:2005zj,2006JCAP...03..007S,2010A&A...523A..28K,Dodelson:2005rf,Schneider:1997ge}.

When studying weak lensing one usually restricts the analysis to small angular separations (or large multipole moments), which is justified by the fact that so far cosmic shear surveys have covered only a limited portion of the sky. However, nearly full-sky surveys are currently under preparation and on large angular scales general relativistic  second-order couplings will become relevant. These corrections are usually neglected because they are suppressed by the ratio between the scale probed and the Hubble scale, but are important on angular scales of the order of the angular diameter distance to the source. These are analogous to the second-order effects from general relativity affecting the CMB bispectrum when at least one of the scales probed is larger than the horizon at recombination \cite{Bartolo:2004ty,Boubekeur:2008kn,Pitrou:2010sn}. In the CMB, these effects have been recently found to be negligible for the contamination of Planck searches for a primordial local signal in the squeezed limit \cite{Creminelli:2011sq,Bartolo:2011wb}. In case of cosmic shear observations, these effects however cannot be a priori neglected. The aim here is to provide an exhaustive evaluation of the impact of these contributions to observations. Those results extend recent studies on the impact of relativistic corrections to the observations of large-scale galaxy clustering in \cite{Yoo:2009au,Bonvin:2011bg,Challinor:2011bk,Baldauf:2011bh,Yoo:2011zc}.

The complete study of all second-order effects in the cosmic shear, including the general relativistic ones, has been recently undertaken in \cite{Bernardeau:2009bm} by solving at second order the Sachs equation  \cite{1961RSPSA.264..309S}, which describes the deformation of the cross-section of a light bundle, mapping galaxy shapes into their angular images. The main results of this paper are reviewed in \sect~\ref{sec:GReffects}.

The goal of the current paper is to compute the bispectrum of the electric part of the cosmic shear from all second-order effects and compare the importance of the general relativistic corrections with the standard couplings. We review the computation of the shear power spectrum in \sect~\ref{sec:PS} where we also present one of the numerical shortcuts we will use throughout the paper, the second-order Limber approximation.
The formal expression of the bispectrum, computed in \sect~\ref{sec:BS}, is determined by two types of non-linear contributions: the dynamical couplings, which depend on the particular metric solution of  Einstein's equations at second order, and the geometrical couplings, so-called because they depend on the geometry of the solution of the Sachs equation at second order. Among the latter we also include the corrections coming from the inhomogeneities of a fixed-redshift source plane.
The calculations of \sect~\ref{sec:BS} involve multiple integrals and complicated manipulations of spherical harmonics and Wigner symbols. (In appendix~\ref{app:spart} one can find useful relations for these calculations.) We are afraid to say that those calculations are extremely lengthy and technical. Impatient readers can skip this section and go directly to \sect~\ref{sec:results}.
In this section we present our results concentrating on the squeezed limit, where one of the multipole  moments $l$ involved in the bispectrum 
is small and we discuss the functional forms of the resulting terms as well as their relative importance. In \sect~\ref{sec:primordial} we compare our results with the bispectrum generated by  primordial non-Gaussianities of the local type and we compute their contamination to a primordial $f_{\rm NL}^{\rm loc}$.
Finally,  in \sect~\ref{sec:conclusion} we conclude and discuss the results of the paper.

\section{Full-sky lensing shear at second order}
\label{sec:GReffects}

\subsection{The shear and spin-$s$ spherical harmonics}

The shear is characterized by a $(2\times 2)$-traceless and symmetric tensor $\gamma_{ab}$, whose components are
defined with respect to a particular choice of axes about each direction on the sky. This tensor describes the
deformation of the image of distant galaxies by the gravitational field of the cosmic structures intervening
between emission and observation. The two independent components of $\gamma_{ab}$, $\gamma_1$ and $\gamma_2$,
can be conveniently combined into a single complex field representing the shear,
\[
\gamma = \gamma_1 + i \gamma_2\;.
\]
Under a right-handed rotation of the axes by an angle $\alpha$ about the line of sight $\hat n$, this complex
field transforms as $\gamma \to e^{i 2 \alpha} \gamma$. Thus, it represents a spin-2 field that can be appropriately
expanded in terms of spin-weighted spherical harmonics, as
\[
\gamma (\hat n)= \sum_{l m} {}_2 a_{lm} \; {}_2 Y_{lm}(\hat n)\;. \label{gamma_exp}
\]
The complex conjugate of $\gamma$, $\gamma^* = \gamma_1 - i \gamma_2$, is then a spin-$(-2)$ field, and can
therefore be expanded as
\[
\gamma^* (\hat n)= \sum_{l m} {}_{-2} a_{lm} \;  {}_{-2}Y_{lm}(\hat n)\;. \label{gamma_*_exp}
\]

One can introduce spin raising and lowering operators that can be used to relate quantities of different spin.
The spin raising operator is denoted as $\spart$ and the spin lowering operator as $\spartb$.
For a general spin-$s$ field ${}_sX$, these are defined as
\begin{align}
\spart\;{}_sX &\equiv
-\sin^s \theta\left(\pa_\theta + i\frac{1}{\sin\theta} \pa_\varphi\right)
(\sin^{-s} \theta)
\; {}_s X \;,
\\
\spartb\;{}_s X &\equiv
-\sin^{-s}\theta\left(\pa_\theta - i\frac{1}{\sin\theta} \pa_\varphi\right) (\sin^{s} \theta)
\; {}_s X \;.
\end{align}
Note that $\spart$ and $\spartb$ commute only when applied to a spin-$0$ quantity. In general one has
\[
\big(\spartb \spart - \spart \spartb \, \big) {}_s X = 2 s \, {}_s X\;. \label{comm_rel}
\]
These operators can be used to obtain spin-0 quantities.
Acting twice with $\spartb$ and $\spart$ respectively on $\gamma$ and $\gamma^*$ in eqs.~\eqref{gamma_exp}
and \eqref{gamma_*_exp}, and using the orthogonality properties of the spherical harmonics one obtains an
expression for the expansion coefficients ${}_2 a_{lm}$ and ${}_{-2} a_{lm}$ as
\begin{align}
{}_2 a_{lm} &= \nico(l,-2) \int d\hat n \ Y_{lm}^*(\hat n)\ \spartb{}^2 \gamma (\hat n)\;, \label{2a}\\
 {}_{-2} a_{lm}& = \nico(l,-2) \int d\hat n \ Y_{lm}^*(\hat n)\ \spart^2 \gamma^* (\hat n)\;, \label{-2a}
\end{align}
where for convenience we have defined $\nico$ as
\[
\nico(l,s) \equiv \sqrt{\frac{(l+s)!}{(l-s)!}}\;. \label{xi}
\]

As for the Cosmic Microwave Background (CMB) polarization, we introduce parity eigenstates, spin-0 quantities, called the ``electric'' and ``magnetic'' parts of the shear, defined as \cite{Kamionkowski:1996ks,Zaldarriaga:1996xe}
\begin{align}
E(\hat n) &\equiv \sum_{lm} a_{E,lm}\ Y_{lm}(\hat n)\;, \\
B(\hat n) &\equiv \sum_{lm} a_{B,lm}\ Y_{lm}(\hat n)\;,
\end{align}
where the coefficients $a_{E,lm}$ and $a_{B,lm}$ are given by
\begin{align}
a_{E,lm} &\equiv -\frac12 ({}_{-2} a_{lm} + {}_{2} a_{lm})\;, \label{E}\\
a_{B,lm} &\equiv -\frac{i}2 ({}_{-2} a_{lm} - {}_{2} a_{lm})\;. \label{B}
\end{align}
Under parity transformation, $E$ and $B$ change as $E \to E$ and $B \to - B$.
Furthermore, in contrast to the shear components $\gamma_1$ and $\gamma_2$, these two spin-0 quantities have the
advantage of being rotationally invariant.

\subsection{The shear up to second order}

We consider a flat FLRW metric background. In \cite{Bernardeau:2009bm} we have computed the shear at second order by using a perturbed metric in the so-called generalized Poisson gauge. In this gauge the second-order metric reads \cite{Bertschinger:1996,Bruni:1996im}
\[
\label{metric2}
\dd s^2=a^2(\eta) \left[ -e^{2\phi}\dd \eta^2+2 \omega_{i}\,\dd \eta\,\dd x^{i}
+ \left(e^{-2\psi} \delta_{ij} +h_{ij} \right) \dd x^{i}\,\dd x^{j}\right]\;,
\]
where the vector component $\omega_i$ is divergenceless, $\partial_i \omega_i=0$, and the tensor component $h_{ij}$
is divergenceless and traceless, $\partial_i h_{ij} = 0 = h_{ii}$.
While the scalar perturbations $\phi$ and $\psi$ contain first- and second-order contributions, since we neglect primordial vector and tensor perturbations, $\omega_i$ and $h_{ij}$ are only second-order quantities.

Note that we have chosen to write the gravitational potentials in the metric in the exponential
form. This choice is convenient for two reasons. First, up to second order, in this form the metric (\ref{metric2}) is conformal to $\dd s^2 = -e^{2(\phi+\psi)} \dd \eta^2 +2 \omega_{i}\,\dd \eta\,\dd x^{i}
+ \left(\delta_{ij} +h_{ij} \right) \dd x^{i}\,\dd x^{j} $ so that the effect of scalar perturbations on
null geodesics will be only through the Weyl potential, defined as
\[
\Psi \equiv (\phi + \psi)/2\;.
\]
The second reason, which will be explained in more details in section~\ref{sec:dyn_coup},
is that with this choice the relativistic second-order contributions to $\phi$ and $\psi$ vanish in the squeezed limit.

In this metric the lensing shear is obtained by solving Sachs equation \cite{1961RSPSA.264..309S},
which describes linear deformations of the infinitesimal cross-section of a light bundle in the optical limit,
and maps galaxy intrinsic shapes into their angular images. At {\em first order} one finds the full-sky expression
of the shear field\footnote{Surprisingly, to our knowledge this full-sky expression was explicitly derived only recently 
in \cite{Bernardeau:2009bm} although it is implicit in \cite{Uzan:2000xv} and in the context of CMB lensing, see for instance 
\cite{2000PhRvD..62d3007H,2003PhRvD..67h3002O}. The form was painstakingly rederived in
 \cite{2010PhRvD..82j3522D} in the context of full-sky cosmic shear observations.}
 \[
\gamma (\hat n)= \int_0^{\chiS} \! d  \chi \frac{\chiS - \chi}{\chiS \chi} \spart^2 \Psi(\chi,\vec x)\;,
\label{shear_first}
\]
where we have conveniently defined $\chi \equiv \eta_0-\eta$, where $\eta_0$ is the conformal time today,
so that $\vec x \equiv \hat n \chi$ is the background photon geodesic. The subscript $S$ in $\chiS$ refers to the source.

At second order, there are four sources of non-linearities \cite{Bernardeau:2009bm}:
\begin{itemize}
\item[1)] The mapping solution of the Sachs equation is linear in the angular deformation, but it is {\em non-linear} in the Weyl potential encountered by the photon from emission to observation. This induces non-linearities even when using the linear part of the metric.
\item[2)] What we observe is actually the reduced shear \cite{Dodelson:2005rf,Schneider:1997ge,2010A&A...523A..28K}, i.e.~the ratio between the anisotropic and isotropic deformations, and there are non-linear corrections introduced when taking this ratio. These will be proportional to the product of the first-order shear and the first-order convergence.
\item[3)] Observationally, we are mapping galaxies located at a given redshift $z_S$. Thus, we expect second-order contributions to the observed shear coming from perturbing $z_S$ in the first-order expression \eqref{shear_first}.
\item[4)] The metric contains second-order terms in the initial conditions. This induces non-linearities even when using the linear solution of the Sachs equation.
\end{itemize}

The first three contributions are independent of the {\em second-order} components of the metric: we will collect them under the name of ``geometrical'' contribution, because they depend on the geometry of the linear mapping, solution of the Sachs equation at second order. On the other hand, the fourth contribution depends only on the linear mapping at first order, but depends on the particular metric solution of the Einstein equations at second order. Thus, it is natural to call it the ``dynamical'' contribution. Note that this separation depends on the particular gauge chosen, hence it is not completely physical. However, for reasons that will appear more clearly in the following, it is a convenient distinction that we adopt hereafter.

\subsubsection{Geometrical couplings}

\label{sec:geom_coup}

Except for the contribution coming from the perturbation of the redshift, which we discuss below, the geometrical couplings depend exclusively on the Weyl potential $\Psi$. As explained above, this is a consequence of the choice of the exponential of $\phi$ and $\psi$ in the form of the metric in eq.~\eqref{metric2}. Some of the geometrical couplings dominate in the large-$l$ limit -- corresponding to small angular scales -- because they contain more operators $\spart$ or $\spartb$. Indeed, in harmonic space to each of such operators is associated an $l$-factor in the angular power spectrum and bispectrum. These dominant terms, that are standard in the literature, can be written as
\[
\label{geom_stand}
\begin{split}
\gamma^{{\rm (stan)}}_{\rm geom}  = &  - \int_0^{\chiS} \! d\chi  \int_0^{\chi} \! d \chi'
\frac{\chiS - \chi}{\chiS \chi} \, \frac{\chi - \chi'}{\chi \chi'}
\Big[ \spart \left(\spart^2 \Psi(\chi,\vec x)   \spartb \Psi(\chi',\vec x')
+ \spartb \spart \Psi(\chi,\vec x)   \spart \Psi(\chi',\vec x')\right)  \\
& \qquad \qquad \qquad \qquad \qquad \qquad \qquad \quad  +  2 \spart \Psi(\chi,\vec x)  \spart \Psi(\chi',\vec x')  \Big] \\
&+\int_0^{\chiS} \! d\chi  \int_0^{\chiS} \! d \chi'   \frac{\chiS - \chi}{\chiS \chi} \,
\frac{\chiS - \chi'}{\chiS \chi'}\spartb \spart \Psi(\chi,\vec x)     \spart^2 \Psi(\chi',\vec x')\;.
\end{split}
\]
The first two lines contain the usual couplings such as the lens-lens correction and the correction to the Born approximation \cite{1997A&A...322....1B,Cooray:2002mj,2003MNRAS.344..857T,Dodelson:2005zj,2006JCAP...03..007S}.\footnote{For comparison with a recent article in the literature on the subject, by rewriting in the first two lines of eq.~\eqref{geom_stand} the spin raising and lowering operators in terms of spatial gradients one recovers the second line of eq.~(8) of \cite{2010A&A...523A..28K}.} The third line is the correction involved in the relation between the shear and the reduced shear, containing the coupling between the first-order convergence and shear \cite{Dodelson:2005rf,Schneider:1997ge}.

If we relax the small-angle approximation and we consider the full sky, there are other terms which become important \cite{Bernardeau:2009bm}. The total geometrical contribution is
\[
\gamma_{\rm geom} = \gamma^{{\rm (stan)}}_{\rm geom}+ \gamma^{{\rm (corr)}}_{\rm geom} + \gamma^{{\rm (z)}}_{\rm geom}\;,
\]
where the standard terms $\gamma^{{\rm (stan)}}_{\rm geom}$ are given in eq.~\eqref{geom_stand}, and the new terms, that we can consider as corrections to the small-angle approximation, are given by
\[
\label{geom_corr}
\begin{split}
\gamma^{{\rm (corr)}}_{\rm geom} =  & - \int_0^{\chiS} \! d\chi \left[ \frac{\chiS - \chi}{\chiS \chi}
\spart^2 \Psi^2(\chi,\vec x) - \frac2{\chiS \chi} \spart^2 \left( \left(\Psi(\chi,\vec x)
-(\chiS-\chi) \dot \Psi(\chi,\vec x)\right) \int_0^{\chi} \! d \chi'  \Psi(\chi',\vec x') \right)  \right]\\
& + 2 \int_0^{\chiS} \! d\chi  \int_0^{\chi} \! d \chi'  \left( \frac{\chiS - \chi}{\chiS \chi^2}  + \frac1{\chiS \chi'}  \right)
\Psi(\chi,\vec x)  \spart^2 \Psi(\chi',\vec x')\\
&-2  \int_0^{\chiS} \! d\chi  \int_0^{\chiS} \! d \chi'  \frac{\chiS - \chi'}{\chiS^2 \chi'}
\Psi(\chi,\vec x)   \spart^2 \Psi(\chi',\vec x')\;.
\end{split}
\]

Note that the second line of  eq.~\eqref{geom_stand} contains only two transverse derivatives and its contribution to the shear is therefore of the same order as the corrections in eq.~\eqref{geom_corr}. Thus, one could be tempted to include this term in the geometrical corrections rather than in the standard contributions. The reason is that classifying terms by counting powers of $\spart$ and $\spartb$ operators leads to ambiguities, as in the full sky  those operators do not commute. It appears that the second line of \eqref{geom_corr} can  be combined with the second term of the first line of eq.~\eqref{geom_stand} to form a {\em single} term containing four spatial gradients,
\[
\spart \spartb \spart \Psi\spart \Psi +2\spart \Psi\spart \Psi=\spartb\spart^2\Psi\spart\Psi\; .
\]
For this reason, we have decided to include this term in the standard contribution.

Finally, the other contribution independent of the second-order components of the metric is the one due to the inhomogeneity of the redshift of the sources. This induces a coupling between the photon redshift $z_S$ and the lens,
\[
\label{geom_z}
\gamma^{{\rm (z)}}_{\rm geom}  =    \frac{1}{\chiS^2 {\cal H}_S} \left(- 2 \int_0^{\chiS} \! d \chi \dot \Psi(\chi,\vec x)+ \phi(\chiS,\vec x_S) -
\hat n \cdot \vec v_S  \right)
\int_0^{\chiS} \! d \chi'  \spart^2 \Psi(\chi',\vec x')\;,
\]
where ${\cal H}$ is the conformal Hubble parameter defined as ${\cal H} \equiv d \ln a / d \eta$,  $\vec v_S$ is the peculiar velocity of the source
and a dot denotes a derivative with respect to $\chi$, $\dot{}=\partial_\chi=-\partial_\eta$.
Note that the inhomogeneity of the source redshift affects the convergence $\kappa$ already at first order in perturbation theory~\cite{2008PhRvD..78l3530B}, but its impact on the shear is only through a coupling with the lens, i.e. of second order.

Since we are interested in ensemble averages over the random field $\phi$ and $\psi$, we need to relate the velocity of the source $\vec v_S$ to the primordial perturbation at first order. We choose to use the linearized Einstein equations to rewrite $\vec v_S$ in terms of the Weyl potential at the source and its time derivative. In doing this we implicitly assume that the source is comoving with the dark matter. At large enough scales this is a legitimate assumption as all matter components are expected to move identically. However, at smaller scales this is clearly an approximation as galaxies typically belong to virialized halos and move independently from the dark matter field.
At first order, the $0j$-component of the Einstein equations reads
\[
\frac2{a^2}\partial_i (\partial_\eta \psi + \Hconf \phi) = - 8 \pi G \rho_m v^i\;.
\]
Thus, using that $\dot{} = -\partial_\eta$, the dark matter velocity at the source reads, at first order,
\[
\vec v_S = - \frac23 \frac{a_S \vec \nabla (- \dot \psi_S + \Hconf_S \phi_S)}{H_0^2 \Omega_{m}}\;, \label{velocity}
\]
where $\Omega_{m}$ is the critical density of matter today, $\Omega_m \equiv 8 \pi G \rho_{m,0}/(3 H_0^2)$, and we have used that $\rho_m = \rho_{m,0}/a^3$.

Note that the lens-lens correction and the correction to the Born approximation in the first line of eq.~\eqref{geom_stand}, can be written as the derivative of a deflection angle. The two shear components associated with these terms are consequently related to the convergence and rotation part of the magnification matrix that contains only two degrees of freedom~\cite{Hirata:2003ka}. On the other hand, the geometrical and redshift corrections cannot be written as the derivative of a deflection angle. Hence, due to these relativistic effects the magnification matrix contains in general four degrees of freedom: the shear $E$-modes are generically different from the convergence and the shear $B$-modes are different from the rotation.

\subsubsection{Dynamical couplings}
\label{sec:dyn_coup}

Before moving to the contribution coming from second-order metric perturbations, i.e.~the dynamical contribution, let us specify the first-order potentials and the initial conditions.
Since we will perform our calculations in a $\Lambda$CDM universe, the anisotropic stress vanishes and the traceless part of the $ij$-components of the Einstein equations imply that $\phi$ and $\psi$ are the same at first order, i.e.~$\phi^{(1)}=\psi^{(1)}=\Psi^{(1)}$. Thus the Weyl potential can be decomposed in Fourier space as
\[
\label{fourier}
\Psi^{(1)}(\chi, \hat n \chi) = \int \frac{d^3 \vec k}{(2 \pi)^3}\,g(\chi)\,T(k)\,\Phi_{\vec k} \ e^{i \vec k \cdot \hat n \chi}\;,
\]
where $T(k)$ is the matter transfer function and $g(\chi)$ is the so-called growth-suppression factor, defined as $g(\chi) \equiv D(\chi)/a(\chi)$, where $D(\chi)$ is the linear growth function. 
Furthermore, $\Phi$ is the primordial potential which represents the initial curvature perturbation generated
during inflation. During matter dominance, on super-horizon scales it is simply proportional to the curvature perturbation on uniform density hypersurfaces,  $\Phi= -(3/5)\zeta$.
For the minimal model of inflation $\zeta$ is approximately Gaussian \cite{Maldacena:2002vr} and in the following we are going to assume that $\Phi$ obeys perfectly Gaussian statistics.

At second order we split both $\phi$ and $\psi$ into a Newtonian part equal for both potentials, denoted by $\phin$, and a relativistic part, denoted respectively by  $\phir$ and $\psir$.  The Newtonian part dominates on small scales; on these scales the relativistic part is suppressed with respect to the Newtonian part by factors of order $(Ha/k)^2$ and becomes relevant only on large scales.
Thus, the Newtonian potential $\phin$ dominates in the small angle approximation and its contribution to the shear has been thoroughly studied in the literature \cite{1997A&A...322....1B,Bernardeau:2002fc,2003A&A...397..405B}. It is obtained from  inserting the second-order Newtonian potential $\phin$
into the first-order expression for the shear \eqref{shear_first}. This yields
\[
\label{dyn_stand}
\gamma^{{\rm (stan)}}_{\rm dyn} =    \int_0^{\chiS} \! d\chi  \frac{\chiS - \chi}{\chiS \chi}
\spart^2 \phin(\chi,\vec x) \;.
\]
The Newtonian potential $\phin$ has been derived in standard Eulerian perturbation theory \cite{Peebles:1982,Bernardeau/etal:2002} and for
$\Lambda$CDM it can be written as 
\[
\label{phin_fourier}
\phin(\chi, \hat n \chi) = \int \frac{d^3 \vec k}{(2 \pi)^3}\,\phin({\vec k},\chi) \, e^{i \vec k \cdot \hat n \chi}\;,
\]
with
\[
\label{phinewt}
\phin(\vec k,\chi)=-\frac{2 g^2a}{3 H_0^2\Omega_{m}}
\frac{T(k_1)T(k_2)}{k^2} k_1^2 k_2^2 F_{2,N}({\vec k_1},{\vec k_2},\chi)\Phi_{\vec k_1}\Phi_{\vec k_2}\; ,
\]
where the kernel $F_{2,N}$ is well approximated by
\[
\label{F2}
F_{2,N}({\vec k_1},{\vec k_2},\chi)=\frac{1}{2}(1+\epsilon)+\frac{\hat k_1 \cdot \hat k_2}{2}\left(\frac{k_1}{k_2}+\frac{k_2}{k_1} \right)+\frac{1}{2}(1-\epsilon)(\hat k_1 \cdot \hat k_2)^2 \; ,
\]
where $\epsilon(\chi)\simeq \frac{3}{7}\left[{\rho_{m}(\chi)}/{\rho_{\rm tot}(\chi)}\right]^{-1/143}$  \cite{Bernardeau/etal:2002,2011JCAP...06..019B}.\footnote{Here we have neglected the contributions to $F_{2,N}$ from the radiation era studied in \cite{Fitzpatrick:2009ci}. These can be easily added to the standard dynamical contribution without changing the other corrections.} Here and below we implicitly assume integration over repeated momenta, i.e.~$\int \frac{d^3 {\vec k_1} d^3 {\vec k_2}}{(2\pi)^3} \delta_D({\vec k}-{\vec k_1}-{\vec k_2})$.
Note that the expression for $F_{2,N}$ above is exact in matter dominance, where $\epsilon=3/7$ and $F_{2,N}$ is time independent. For $\Lambda$CDM the time dependence of $\phin$ cannot be factorized out in the growth-suppression factor $g^2(\chi)$ in front of the integral but $F_{2,N}$ is mildly time dependent.  The approximation above to characterize this time dependence reproduces very well the one given
in \cite{Fitzpatrick:2009ci}.

On large scales the relativistic parts of the scalar potentials  $\phir$ and $\psir$ become important. Also the vector and the tensor modes in the metric, respectively $\omega_i$ and $h_{ij}$, generated at second order are relativistic corrections to the metric of the same order in $aH/k$ as the relativistic potentials $\phir$ and $\psir$. As they are intrinsically second order, they simply enter linearly in the expression for the shear, yielding
\[
\begin{split}
\gamma^{{\rm (corr)}}_{\rm dyn} & =   \int_0^{\chiS} \! d\chi  \left[  \frac{\chiS - \chi}{\chiS \chi}
\spart^2  \left( \frac12 \big( \phir (\chi,\vec x) + \psir (\chi,\vec x) \big)- \frac12 \omega_r (\chi,\vec x)-\frac14  h_{rr} (\chi,\vec x)\right) \right. \\ &
  \qquad \qquad \quad  - \left. \frac1{2 \chi}  \spart \left( {}_1\omega (\chi,\vec x) +  {}_1h_r (\chi,\vec x) \right)\right] - \frac14 {}_2 h(\chiS,\vec x_S)\;,\label{gamma_corr_dyn}
\end{split}
\]
where we have defined  $\omega_r  \equiv \hat n^i \omega_i$ and $h_{rr}  \equiv  \hat n^i \hat n^j  h_{ij}$,
the spin-1 part of $\omega_i$ and $h_{ij}$ respectively as ${}_1 \omega  \equiv \hat e_+^i \omega_i$ and
${}_1 h_{r}  \equiv \hat e_+^i \hat n^j  h_{ij}$, and the spin-2 part of $h_{ij}$ as ${}_2 h \equiv \hat e_+^i \hat e_+^j h_{ij}  $. 
Here $\hat e_+\equiv \hat e_\theta+ i \hat e_\varphi$ where $\hat e_\theta$ and $\hat e_\varphi$ are unit vectors orthogonal to the photon propagation $\hat n$
and to each other. 
The term containing ${}_2 h$ in the second line of this equation is a boundary term induced by the spin-2 part of the tensor modes,
which account for the distortion of the shape at the source.
Putting together the dynamical contributions of eqs.~\eqref{dyn_stand} and \eqref{gamma_corr_dyn} we obtain \cite{Bernardeau:2009bm}
\[
\gamma_{\rm dyn}^{\phantom{()}} = \gamma^{{\rm (stan)}}_{\rm dyn}+ \gamma^{{\rm (corr)}}_{\rm dyn}\;.
\]

Let us give here the expressions for the relativistic components of the metric to be put in eq.~\eqref{gamma_corr_dyn}. For a $\Lambda$CDM universe, these have been computed in \cite{Bartolo:2005kv} and their expressions are quite involved. However, as we will see in section~\ref{sec:results}, the effect of the dynamical relativistic terms on the shear bispectrum is extremely small. For this reason we can estimate their contribution by simply using the expressions for the relativistic components in matter dominance, where it is possible to factorize their time and momentum dependence. Indeed, in this case these metric components can be written, in Fourier space, as \cite{Boubekeur:2008kn,Boubekeur:2009}

\begin{align}
\phir(\vec k,\chi) & =  \frac{T(k_1) T(k_2)}{k^2} \left[ \vec k_1 \cdot \vec k_2 -
3 {(\hat k \cdot \vec k_1)(\hat k \cdot \vec k_2)} \right] \Phi_{\vec k_1} \Phi_{\vec k_2}\;, \label{phi2} \\
\psir(\vec k,\chi) & = - \frac23 \phir(\vec k,\chi)\;,\label{psi2}\\
\omega_i (\vec k,\chi)&=   - \frac{i 4 a^{1/2}}{3 H_0 } \frac{T(k_1) T(k_2)}{k^2}
\left[ k_1^2 k^i_2 + k_2^2 k^i_1  - \hat k^i \left( k_1^2 (\hat k \cdot \vec k_2) + k_2^2 (\hat k \cdot \vec k_1) \right) \right]
\Phi_{\vec k_1} \Phi_{\vec k_2}\;, \label{omega_four}\\
\begin{split}
{h}_{ij} (\vec k,\chi) & =   \frac{10}{3}
\left[ 1 - 3 {j_1(k\eta)}/({k \eta}) \right]
\frac{T(k_1) T(k_2)}{k^4} \left[ \big( (\vec k_1 \cdot \vec k_2)^2 -k_1^2 \, k_2^2 \big) (\delta^{ij}+ \hat k^i \hat k^j) \right.
\\&+ \left.2 \,k_1^2 \, k_2^i\, k_2^j + 2 \,k_2^2 \,k_1^i \,k_1^j - 4\, k_1^i \,k_2^j \, \vec k_1 \cdot \vec k_2 \right]
\Phi_{\vec k_1} \Phi_{\vec k_2} \label{gamma} \, .
\end{split}
\end{align}
In the expression for tensor modes, the spherical Bessel function $j_1(x)$ is given by $j_1(x)=\sin(x)/x^2-\cos(x)/x$.

One can check that this metric solution satisfies Gaussian primordial initial conditions. Indeed, on large scales and at non-linear order the gauge transformation from the $\zeta$ gauge \cite{Maldacena:2002vr} to Poisson gauge \eqref{metric2} is given by $  \zeta = -  \psi-\frac23 \phi$ in matter dominance. Thus, eq.~\eqref{psi2} implies Gaussian initial conditions $\zeta^{(2)}=0$.
Furthermore, we note that in the squeezed limit, i.e.~when one of the two modes $\vec k_1$ and $\vec k_2$ is much larger than the other, the second-order relativistic components of the metric given above go to zero. This is another advantage of using the metric \eqref{metric2} with $\phi$ and $\psi$ in the exponentials.

To summarize, the full-sky observed shear up to second order is given by the sum of five contributions, three geometrical and two dynamical \cite{Bernardeau:2009bm}:
\[
\gamma = \gamma^{\rm (stan)}_{\rm geom}+ \gamma^{\rm (corr)}_{\rm geom} +  \gamma^{\rm (z)}_{\rm geom} + \gamma^{\rm (stan)}_{\rm dyn} +\gamma^{\rm (corr)}_{\rm dyn} \;.
\]
Their expressions are reported, respectively, in eqs.~\eqref{geom_stand}, \eqref{geom_corr}, \eqref{geom_z}, \eqref{dyn_stand} and \eqref{gamma_corr_dyn}.

\section{Angular power spectrum}
\label{sec:PS}

The aim of this section and of the following one is to give the explicit forms of the spectrum and bispectrum of the shear field. More specifically, we are interested in the spectra of the electric part of the shear field. This section focuses on the power spectrum calculation. To present the formalism it is convenient  to introduce the lensing potential $\LP$ through the relation,
\[
\label{Psi_lens_def}
\gamma(\hat n) \equiv \spart{}^2  \LP (\hat n)  \;,
\]
where the shear correlation properties are entirely  encoded into the statistical properties of $\LP(\hat n)$.

\subsection{From metric to shear spectrum}

To compute the lensing power spectrum we just need the shear at {\em linear order} in the Weyl potential. In this case, using the expression for the shear at first order, eq.~\eqref{shear_first}, eq.~\eqref{Psi_lens_def} reduces to the usual definition of the lensing potential given in the literature (see e.g.~\cite{Lewis:2006fu}),
\[
\LP (\hat n) = \int_0^{\chiS} \! d  \chi \frac{\chiS - \chi}{\chiS \chi}  \Psi(\chi,\hat n \chi)\;.
\]
As we will see below when computing the bispectrum, this expression is inappropriate to describe the lensing potential at {\em second order}, while eq.~\eqref{Psi_lens_def} remains valid.

The lensing potential $\LP$ is a spin-$0$ operator. Thus, using the commutation relation \eqref{comm_rel} one shows that
\[
\begin{split}
\spartb{}^2 \gamma = \spartb{}^2 \! \spart^2 \LP &= \spart^2 \spartb{}^2 \LP \\
&= \spart^2 \spartb{}^2 \LP^* = \spart^2 \gamma^*\;,
\end{split}
\]
where to write the second line we have used that $\LP$ is {\em real} at first order. From eqs.~\eqref{2a} and \eqref{-2a}, this equation implies that ${}_2a_{lm} = {}_{-2}a_{lm}$ and, using the definition
of $a_{E,lm}$, eq.~\eqref{E}, that
\[
a_{E,lm} = - \nico(l,2) \int d\hat n \; Y_{lm}^*(\hat n) \LP (\hat n)\;. \label{a_of_Psi}
\]
Note that $a_{B,lm}$ vanishes at first order (see eq.~\eqref{B}). The angular power spectrum $C^E_l$ of the electric part of the shear, $E(\hat n)$, is defined from the 2-point function,
\[
\langle a_{E,lm}^{\phantom{*}} a^*_{E,l'm'} \rangle \equiv C^E_l \delta_{l l'} \delta_{m m'}\;. \label{C_l}
\]
As at this order the angular power spectrum of the magnetic part of the shear is zero, $C_l^B = 0$, there is no ambiguity in simply setting $C_l \equiv C^E_l$ in what follows.

We then use the Fourier mode decomposition of eq.~\eqref{fourier} for the Weyl potential and expand the plane wave in spherical
harmonics, i.e.
\[
\label{exp_spher}
e^{i \vec k \cdot \hat n \chi} = 4 \pi \sum_{lm}\, i^l\, j_l(k \chi)\, Y_{lm}(\hat n)\, Y^*_{lm}(\hat k)\;.
\]
Combining eq.~\eqref{exp_spher} with eq.~\eqref{Psi_lens_def} and \eqref{a_of_Psi}, and integrating in $d \hat n$ with the orthogonality  relation of the spherical harmonics, we obtain an expression for $a_{E,lm}$ as a function of the primordial potential $\Phi_{\vec k}$,
\[
a_{E,lm} = -   i^l \nico(l,2) \int_0^{\chiS} \! d\chi\, W (\chiS, \chi)
\int \frac{d^3 \vec k}{2 \pi^2}\,j_l(k \chi)\,Y^*_{lm}(\hat k)\,T(k)\,\Phi_{\vec k}  \;, \label{aE_lin}
\]
where we have defined the window function $W$ as
\[
W(\chi',\chi) \equiv \frac{\chi'-\chi}{\chi'\chi} g(\chi) \;.
\]

Finally, using eq.~\eqref{aE_lin} in the definition of the $C_l$, eq.~\eqref{C_l}, one obtains
\[
C_l =\nico^2(l,2)
\int_0^{\chiS}\! d\chi \,W(\chiS,\chi)
\int_0^{\chiS}\! d\chi'\,W(\chiS,\chi')
\int \frac{2 k^2 dk}{\pi}\,T^2(k)\,P(k)
\,j_l(k\chi)\,j_l(k \chi')\;, \label{C_l_2}
\]
which can be rewritten in a more compact form as
\[
C_l =\nico^2(l,2)
\int \frac{2 k^2 dk}{\pi}\, P(k)\,T^2(k)
\left[\int_0^{\chiS}\! d\chi W(\chiS,\chi)\,j_l(k \chi)\right]^2\;.
\label{C_l_3}
\]
Here $P(k)$ is the power spectrum for the primordial perturbation $\Phi$, defined by
\[
\langle \Phi_{\vec k} \Phi_{\vec k'} \rangle \equiv (2\pi)^3 \delta(\vec k + \vec k') P(k)\;. \label{spectrum_Phi}
\]
For simplicity in our calculations we will only consider a scale invariant spectrum,
\[
P(k) = {A_{\Phi}}{k^{-3}}\;, \label{Pk}
\]
although all the treatment can be easily extended to a non-zero tilt. In the following we will consider a $\Lambda$CDM cosmology
with $\Omega_m=0.3$ today and a Hubble parameter $h=0.65$. Furthermore, to ease the numerical treatment, for $T(k)$ we will use the BBKS fitting formula~\cite{Bardeen:1985tr}.

\subsection{Numerical integrations: the Limber approximation}

In practice, integrating eq.~\eqref{C_l_3} is numerically involved and time consuming. To compute the $C_l$ it is common to employ the so-called Limber approximation \cite{1992ApJ...388..272K}, which is valid for large $l$. This is based on the fact that the ordinary Bessel function $J_\nu(x)$, related to the spherical Bessel  function $j_l(x)$ by
\[
j_l(x) = \sqrt{\frac{\pi}{2x}} J_\nu(x)\;, \qquad \nu \equiv l+1/2\;,
\]
grows monotonically from zero at $x=0$ to $x \simeq \nu$ and then rapidly oscillates. For large $l$ an integral of an arbitrary function multiplied by a Bessel function can be approximated by
\[
\int dx f(x) J_\nu (x) =  f(\nu) + {\cal O} (1/{\nu^2})\;, \label{limber1}
\]
which can be written in the following form:
\[\label{limber1_dubble_jl}
\int\frac{2k^2\,dk}{\pi}f(k)\,j_l(k\chi)\,j_l(k\chi')
=\frac{\delta(\chi-\chi')}{\chi^2}f(\nu/\chi)
 \left[1 + {\cal O}(1/\nu^2) \right]\;.
\]
Using this approximation one obtains the Limber-approximated $C_l$ as
\[
C_l = \nico^2(l,2)
\int_0^{\chiS} d\chi \frac{W^2(\chiS,\chi)}{\chi^2}\,T^2(\nu/\chi)\,P(\nu/\chi) \left[1 + {\cal O}(1/\nu^2) \right] \;.
\label{Cl_limber}
\]
This expression approximates the exact $C_l$, obtained by integrating the full expression in eq.~\eqref{C_l_3}, to better than
$1\%$ for $l \gtrsim 8$.

For small $l$ this approximation fails. Since we are interested in studying the shear on the full sky and correlate fields lying at different $\chi$ along the line of sight, we need to go beyond the Limber approximation~\eqref{limber1}. As shown in \cite{LoVerde:2008re}, an integral of an arbitrary function multiplied by a Bessel function admits the series representation
\[
\label{Limber2_series}
\begin{split}
\int dx f(x) J_\nu (x) & = \left[ f(x) - \frac12 \frac{x^2}{\nu^2}f''(x) - \frac{1}{6} \frac{x^3}{\nu^2} f'''(\nu)  \right]_{x=\nu}  + {\cal O} (1/{\nu^4})  \;.
\end{split}
\]
This can be used to go at {\em second order} in the Limber approximation -- i.e.~to include corrections of order $1/\nu^2$, such as the second and third terms inside the bracket in the equation above -- and to considerably improve the Limber approximation. The expression for the angular power spectrum at second order in the Limber approximation is given in appendix~\ref{app:limber}, eq.~\eqref{Cl_limber2}.

\begin{figure}[t]\centering
\includegraphics[width=0.6\textwidth]{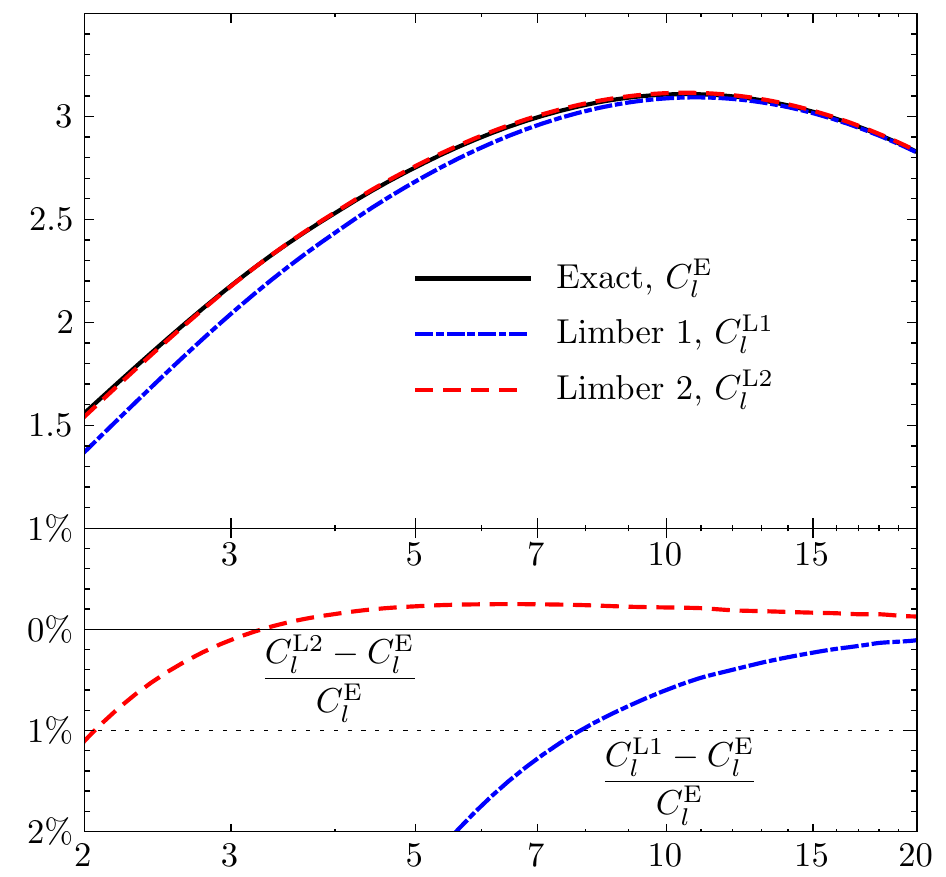}
\caption{Angular power spectrum computed using the exact expression eq.~\eqref{C_l_3} ({\em solid black line}), using the Limber approximation at first order as in eq.~\eqref{Cl_limber} ({\em dotted-dashed blue line}) and at second order as described in appendix~\ref{app:limber}, eq.~\eqref{Cl_limber2} ({\em dashed red line}). The redshift of the source is $z_S=1$. In the lower panel we have plotted the error made using the first- and second-order Limber approximation. The error made using the second-order approximation is smaller than $1\%$ even at very small $l$.}
\label{fig:Cl}
\end{figure}

In fig.~\ref{fig:Cl} we show the angular power spectrum computed using the exact expression eq.~\eqref{C_l_3} compared with the Limber approximation at first and second order. By going at second order in the Limber approximation, one obtains an agreement with the exact expression to better than $1\%$ for $l \gtrsim 2$.

In practice and for the calculations of the spectra and bispectra, except when otherwise stated, we will employ this improved approximation.

\clearpage
\section{Angular bispectrum}
\label{sec:BS}

\subsection{The $E$-mode reduced bispectrum}

At second order, the lensing potential defined in eq.~\eqref{Psi_lens_def} is in general a complex quantity. Consequently
${}_2a_{lm}$ and ${}_{-2}a_{lm}$ are not equal anymore. This implies that the second-order shear will generically have both $E$-modes and $B$-modes.

As in the first-order case, we can express $a_{E,lm}$ and $a_{B,lm}$ as a function of the lensing potential.
From eqs.~\eqref{2a} and \eqref{-2a} and using the definition of $a_{E,lm}$ and $a_{B,lm}$, respectively eqs.~\eqref{E} and \eqref{B}, one finds
\begin{align}
a_{E,lm} &= - \nico(l, - 2) \int d\hat n \; Y_{lm}^*(\hat n) \Re[ \spartb{}^2 \gamma (\hat n)]= - \nico(l,2) \int d\hat n \; Y_{lm}^*(\hat n) \Re[ \LP (\hat n)]  \;, \label{aE_of_Psi}  \\
a_{B,lm} &= - \nico(l, - 2) \int d\hat n \; Y_{lm}^*(\hat n) \Im[ \spartb{}^2 \gamma (\hat n)]  = - \nico(l,2) \int d\hat n \; Y_{lm}^*(\hat n) \Im[ \LP (\hat n)] \;. \label{aB_of_Psi}
\end{align}
These expressions relate the electric and magnetic part of the shear respectively to {\em real} and {\em imaginary} parts of the lensing potentials.

The angular lensing bispectrum $B^{XYZ}_{l_1 l_2 l_3}$ is defined by
\[
\langle a_{X,l_1 m_1} a_{Y,l_2 m_2} a_{Z,l_3 m_3}\rangle \equiv
\begin{pmatrix}l_1 & l_2 & l_3\\m_1 & m_2 & m_3\end{pmatrix}
B^{XYZ}_{l_1l_2l_3}\;,
\]
where $X$, $Y$ and $Z$ can be either the electric or magnetic parts of the shear, $E$ and $B$, and
\[
\begin{pmatrix}l_1 & l_2 & l_3\\m_1 & m_2 & m_3\end{pmatrix}\;
\]
is the Wigner $3$-$j$ symbol.
As $B$ vanishes at first order, only expectation values containing three $a_{E,lm}$ or two $a_{E,lm}$ and one $a_{B,lm}$ will be non-zero. For parity reasons, the bispectrum involving three $a_{E,lm}$ will be zero for $l_1+l_2+l_3=$ odd, while the one involving two $a_{E,lm}$ and one $a_{B,lm}$ will be zero for $l_1+l_2+l_3=$ even.

In the following we concentrate on computing the bispectrum of the electric part, which will be simply denoted as $B_{l_1l_2l_3}$, defined as
\[
\langle a_{E,l_1 m_1} a_{E,l_2 m_2} a_{E,l_3 m_3}\rangle \equiv
\begin{pmatrix}l_1 & l_2 & l_3\\m_1 & m_2 & m_3\end{pmatrix}
B_{l_1l_2l_3}\;.
\]
Following the standard CMB notation in the literature \cite{Komatsu:2001rj}, we will concentrate on the reduced bispectrum $b_{l_1l_2l_3}$ defined as
\[
\langle a_{E,l_1 m_1} a_{E,l_2 m_2} a_{E,l_3 m_3}\rangle \equiv {\cal G}_{l_1 l_2 l_3}^{m_1 m_2 m_3} \; b_{l_1l_2l_3}\;, \label{reduce_b}
\]
where
\[
\begin{split}
{\cal G}_{l_1 l_2 l_3}^{m_1 m_2 m_3} & \equiv \int d \hat n \; Y_{l_1 m_1} (\hat n)Y_{l_2 m_2} (\hat n)Y_{l_3 m_3}(\hat n)\\
& =\sqrt{\frac{(2l_1+1)(2l_2+1)(2l_3+1)}{4\pi}}
\begin{pmatrix}l_1 & l_2 & l_3\\0 & 0 & 0\end{pmatrix}
\begin{pmatrix}l_1 & l_2 & l_3\\m_1 & m_2 & m_3\end{pmatrix}
\label{Gaunt}
\end{split}
\]
is the Gaunt integral, which characterizes the angular dependence of the bispectrum. It naturally appears due to translational and rotational invariance on the sky, and enforces $m_1$, $m_2$, and $m_3$ to sum up to zero, and $l_1$, $l_2$ and $l_3$ to satisfy the triangle inequality.

To plot our results,  it is convenient to introduce the weighted bispectrum, defined as
\[
\label{weighted}
\hat b_{l_1l_2l_3} \equiv  \frac{b_{l_1l_2l_3}}{C_{l_1} C_{l_2}+C_{l_1} C_{l_3}+C_{l_2} C_{l_3}}\;.
\]
Note that the weighted bispectrum is independent of the power spectrum normalization $A_\Phi$ in eq.~\eqref{Pk}.

\subsection{Geometrical contribution}
\label{sec:geometrical}

We start by computing the contribution of the geometrical terms to the bispectrum.
In order to do this, let us rewrite eq.~\eqref{aE_of_Psi} by moving the operator $\spartb{}^2$ in front of $Y_{lm}^*(\hat n)$ by integrating by parts twice, and by using eqs.~\eqref{def_sw_1} and \eqref{def_sw_2} in appendix~\ref{app:spart}. After these operations,  the coefficient $a_{E,lm}$ becomes
\[
a_{E,lm} = (-1)^{m+1} \int d\hat n  \big[  {}_{-2}Y_{lm}^*(\hat n) \; \gamma (\hat n) + {}_{2}Y_{lm}^*(\hat n) \; \gamma^* (\hat n) \big]\;.
\]

Now, let us initially consider only the standard geometrical contribution, eq.~\eqref{geom_stand}.
Plugging this contribution in the above equation, replacing the Weyl potentials by their Fourier components using eq.~\eqref{fourier} and employing the plane wave expansion \eqref{exp_spher}, this can be rewritten as
\[
\begin{split}
\label{sec_ord_gs}
[a_{\rm geom}^{\rm (stan)}]_{E,lm} =
&\ (-1)^{m+1} \!\!\!\!\! \sum_{l_1 l_2 m_1 m_2}  \!\!\!\!\! i^{l_1+l_2}\!\! \int d\hat n\
	{}_{-2}Y_{lm}^*(\hat n)
\\
&\ \Bigg\{-\Bigg[
	\spart \left(\spart^2 Y_{l_1m_1}^*(\hat n) \spartb Y_{l_2m_2}^*(\hat n)
		+\spartb \spart Y_{l_1m_1}^*(\hat n) \spart Y_{l_2m_2}^*(\hat n) \right)
	+2 \spart Y_{l_1m_1}^*(\hat n)  \spart Y_{l_2m_2}^*(\hat n) \Bigg]
\\
&\ \qquad\qquad \times
\int_0^{\chiS}\!d\chi \int_0^{\chi}\!d\chi'W(\chiS,\chi) W(\chi,\chi') {\cal M}(\chi,\chi')
\\
&\ \phantom{\Bigg\{} +\spartb \spart Y_{l_1m_1}^*(\hat n) Y_{l_2m_2}^*(\hat n) \int_0^{\chiS} \! d\chi
		\int_0^{\chiS} \! d \chi' W(\chiS,\chi) W(\chiS,\chi') {\cal M}(\chi,\chi') \Bigg\}
\\
&\ + \big[ \spart \leftrightarrow \spartb\ \big] \;,
\end{split}
\]
where ${\cal M}$ is a function of $\chi$ and $\chi'$ defined as
\[
{\cal M} (\chi,\chi') \equiv \int \frac{d^3k_1}{2 \pi^2} \frac{d^3k_2}{2 \pi^2} T(k_1) T(k_2)   j_{l_1}( k_1 \chi) j_{l_2}( k_2 \chi')   Y_{l_1m_1} (\hat k_1) Y_{l_ 2 m_2} (\hat k_2)  \Phi_{\vec k_1} \Phi_{\vec k_2}\;.
\]
In the fifth line  $\big[\spart \leftrightarrow \spartb\ \big]$  stands for the first four lines (on the right-hand side of the equality) after replacement of the spherical harmonic ${}_{-2}Y_{lm}^*$ by ${}_{2}Y_{lm}^*$ and of all the spatial gradients by their complex conjugates.

We have thus factorized out the projection operators from the time and momentum integrals. We can now make use of the properties of the spin-weighted spherical harmonics, described in appendix~\ref{app:spart}. It is then straightforward to rewrite the integrals over the angle $\hat n$ in terms of the 3-$j$ symbols using
\[
\int d \hat n \; {}_{s_1}Y_{l_1 m_1} (\hat n){}_{s_2}Y_{l_2 m_2} (\hat n){}_{s_3}Y_{l_3 m_3}(\hat n)
 =\sqrt{\frac{(2l_1+1)(2l_2+1)(2l_3+1)}{4\pi}}
 \begin{pmatrix}l_1 & l_2 & l_3\\s_1 & s_2 & s_3\end{pmatrix}
 \begin{pmatrix}l_1 & l_2 & l_3\\m_1 & m_2 & m_3\end{pmatrix}
 \label{9j} \;.
\]
Note that the 3-$j$ symbol enforces $s_1 +s_2 +s_3=0$.  The fifth line of eq.~\eqref{sec_ord_gs}, $\big[\spart \leftrightarrow \spartb\ \big]$, gives an identical expression as for the first four lines after sign change of the spin indices $s_1$, $s_2$ and $s_3$. Using the following useful property of the 3-$j$ symbols,
\[
\begin{pmatrix}
l_1&l_2&l_3\\
s_1&s_2&s_3
\end{pmatrix}
=
(-1)^{l_1+l_2+l_3}
\begin{pmatrix}
l_1&l_2&l_3\\
-s_1&-s_2&-s_3
\end{pmatrix}
\;,
\]
one finds that for $l_1+l_2+l_3$ even this contribution is identical to the one of the first four lines while for $l_1+l_2+l_3$ odd it is opposite so that all the contributions exactly cancel. Thus, as expected $a_{E,lm}$ vanishes for $l_1+l_2+l_3$ odd.

The bispectrum is then obtained by correlating the second-order $[a^{\rm (stan)}_{\rm geom}]_{E, l m}$ coefficient in eq.~\eqref{sec_ord_gs} with the product of two first-order $a_{E, l m}$, whose expression is given in eq.~\eqref{aE_lin}. After taking the expectation value over the primordial perturbations using Wick's theorem and the definition of the primordial power spectrum, eq.~\eqref{spectrum_Phi}, we obtain for the standard geometrical contribution, for $l_1+l_2+l_3$ even,
\[
\label{bgeomstand}
\begin{split}
[b^{\rm (stan)}_{\rm geom}]_{l_1 l_2 l_3}   = \
& - \; \nico(l_1,2)\nico(l_2,2)
\begin{pmatrix}l_1 & l_2 & l_3\\0 & 0 & 0\end{pmatrix}^{-1}
\\
&\quad\times\left[ \nico(l_1,3)\nico(l_2,1)
\begin{pmatrix}l_1 & l_2 & l_3\\3 &-1 & -2\end{pmatrix}
+\nico^2(l_2,1)\nico(l_1,2)
\begin{pmatrix}l_1 & l_2 & l_3\\2 &0 & -2\end{pmatrix}  \right.
\\
&\quad\phantom{\times\Big[}+ \left. \nico^2(l_1,1)\nico(l_2,2)
\begin{pmatrix}l_1 & l_2 & l_3\\0 &2 & -2\end{pmatrix}
+\nico(l_1,1)\nico(l_2,1)(l_1+2)(l_1-1)
\begin{pmatrix}l_1 & l_2 & l_3\\1 &1 & -2\end{pmatrix}
\right]
\\
&\quad\times \int_0^{\chiS}\!\!\! d \chi_1  d\chi_2   d\chi_3 \int_0^{\chi_3} \!\!\! d \chi'  W(\chiS,\chi_1) W(\chiS,\chi_2) W(\chiS,\chi_3) W(\chi_3,\chi')  {\cal C} (\chi_1,\chi_3;\chi_2,\chi')
\\
&+ \nico^2(l_1,1) \nico(l_1,2) \nico^2(l_2,2)
\begin{pmatrix}l_1 & l_2 & l_3\\0 &2 & -2\end{pmatrix}
\begin{pmatrix}l_1 & l_2 & l_3\\0 & 0 & 0\end{pmatrix}^{-1}
\\
&\quad\times \int_0^{\chiS} d \chi_1  d\chi_2   d\chi_3 \int_0^{\chiS} \! d \chi'  W(\chiS,\chi_1) W(\chiS,\chi_2) W(\chiS,\chi_3) W(\chiS,\chi')  {\cal C} (\chi_1,\chi_3;\chi_2,\chi')
\\
+&\ 5 \ {\rm perms} \;,
\end{split}
\]
where
\[
{\cal C} (\chi,\chi';\chi'',\chi''') \equiv  \int \frac{2 k_1^2 dk_1}{\pi} \frac{2 k_2^2 dk_2}{\pi}
T^2(k_1) T^2(k_2) j_{l_1}(k_1 \chi) j_{l_1}(k_1 \chi') j_{l_2}(k_2 \chi'') j_{l_2}(k_2 \chi''') P(k_1) P(k_2)\; .
\]

The bispectrum from the corrections to the geometrical contribution can be straightforwardly computed from eq.~\eqref{geom_corr}, analogously to the calculation for the standard contribution. For $l_1+l_2+l_3$ even we obtain
\[
\label{bgeomcorr}
\begin{split}
[b^{\rm (corr)}_{\rm geom}]_{l_1 l_2 l_3}   =
&\ \nico(l_1,2) \nico(l_2,2) \nico(l_3,2)
\int_0^{\chiS} d \chi_1  d\chi_2   d\chi_3 W(\chiS,\chi_1) W(\chiS,\chi_2)
\\
&\quad\times \bigg[  W(\chiS,\chi_3) g(\chi_3)  {\cal C} (\chi_1,\chi_3;\chi_2,\chi_3)
\\
&\qquad\qquad+ \frac{2}{\chiS\chi_3}  \bigg(  -g(\chi_3) + (\chiS - \chi_3) g'(\chi_3)  \bigg) \int_0^{\chi_3} d\chi' g(\chi') {\cal C} (\chi_1,\chi_3;\chi_2,\chi') \bigg]
\\
&\ -2 \nico(l_1,2) \nico^2(l_2,2)
\begin{pmatrix}l_1 & l_2 & l_3\\0 & 2 & -2\end{pmatrix}
\begin{pmatrix}l_1 & l_2 & l_3\\0 & 0 & 0\end{pmatrix}^{-1}
\int_0^{\chiS} d \chi_1  d\chi_2   d\chi_3  W(\chiS,\chi_1) W(\chiS,\chi_2)
\\
&\quad \times   \bigg[ \int_0^{\chi_3} \! d \chi'   \left( \frac{W(\chiS , \chi_3)}{\chi_3}  + \frac{g(\chi_3)}{\chiS \chi'}  \right)  g(\chi')  {\cal C} (\chi_1,\chi_3;\chi_2,\chi')
\\
& \qquad\qquad -\frac{g(\chi_3)}{\chiS}\int_0^{\chiS} d \chi' W(\chiS,\chi')  {\cal C} (\chi_1,\chi_3;\chi_2,\chi')  \bigg]
\\
+& \ 5 \ {\rm perms}\;.
\end{split}
\]

The calculation of the bispectrum from the redshift correction follows from eq.~\eqref{geom_z} and proceeds as above. The only complication resides in the spatial gradient in the velocity term, eq.~\eqref{velocity}, projected along the line of sight. Using $\hat n \cdot \vec \nabla e^{i \vec k \cdot \hat n \chi} = \partial_\chi e^{i \vec k \cdot \hat n \chi} $, this simply introduces a derivative with respect to $\chi$ of the spherical Bessel function in the expansion of plane waves into spherical harmonics. In this case the bispectrum reads
\[
\label{bgeomz}
\begin{split}
[b^{\rm (z)}_{\rm geom}]_{l_1 l_2 l_3}   = \ & \; \nico^2(l_1,2)\nico(l_2,2)   \begin{pmatrix}l_1 & l_2 & l_3\\2 & 0 & -2\end{pmatrix} 
\begin{pmatrix}l_1 & l_2 & l_3\\0 & 0 & 0\end{pmatrix}^{-1}
\\
&\times  \int_0^{\chiS}d\chi_1d\chi_2 d\chi_3W(\chiS,\chi_1)W(\chiS,\chi_2)g(\chi_3)
\\
&\qquad\Bigg \{
\frac{1}{\Hconf_S\chiS^2}  \Bigg[2\int_0^{\chiS}d\chi'g'(\chi')  {\cal C}(\chi_1,\chi_3;\chi_2,\chi')
- g(\chiS) {\cal C}(\chi_1,\chi_3;\chi_2,\chiS) \Bigg]
\\
&\qquad\qquad +\frac{2 a_S }{3H_0^2\Omega_m\chiS^3}
\left[g(\chiS)-\frac{g'(\chiS)}{\Hconf_S}\right]  {\cal D}(\chi_1,\chi_3;\chi_2,\chiS)
\Bigg \}
\\
+&\  5 \ {\rm perms}\;,
\end{split}
\]
where
\[
\label{jlp}
{\cal D} (\chi,\chi';\chi'',\chi''') \equiv \int \frac{2 k_1^2dk_1}{\pi} \frac{2 k_2^2 dk_2}{\pi} T^2(k_1) T^2(k_2) j_{l_1}(k_1 \chi) j_{l_1}(k_1 \chi') j_{l_2}(k_2 \chi'')\frac{\partial j_{l_2}(k_2 \chi''')}{\partial \ln \chi'''} P(k_1) P(k_2)\;.
\]

\subsection{Dynamical contribution}
\label{sec:dynamical}

We compute here the bispectrum from the dynamical terms. Let us start by the Newtonian contribution, eq.~\eqref{dyn_stand}. For this contribution the lensing potential is real and reads, using eq.~\eqref{Psi_lens_def},
\[
\LP_{\rm dyn}^{\rm (stan)} = \int_0^{\chiS} d \chi \frac{\chiS - \chi}{\chiS \chi} \phi_N^{(2)} (\chi, \vec x)\;.
\]
Plugging this expression with \eqref{phinewt} into eq.~\eqref{aE_of_Psi}, expanding the plane wave in spherical harmonics and integrating over the angle $\hat n$ using the orthogonality of the spherical harmonics we find
\[
\begin{split}
[a_{\rm dyn}^{\rm (stan)}]_{E,l m }=
&\ i^{l}  \nico(l,2)\frac{2}{3 H_0^2\Omega_m}
\int_0^{\chiS}d\chi W(\chiS,\chi)g(\chi)a(\chi)
\int \frac{d^3 {\vec k_3}}{2\pi^2}Y^*_{l m }(\hat k_3)j_{l}(k_3 \chi)
\\
\times &\int \frac{d^3 {\vec k_1} d^3 {\vec k_2}}{(2\pi)^3}
\delta({\vec k_3}-{\vec k_1}-{\vec k_2})
T(k_1)T(k_2)
\frac{k_1^2 k_2^2}{k_3^2}F_{2,N}({\vec k_1},{\vec k_2},\chi)
\Phi_{\vec k_1}\Phi_{\vec k_2} \; ,
\end{split}
\]
where $F_{2,N}$ is given in eq.~\eqref{F2}.

Using eq.~\eqref{aE_lin}, the expectation value of three $a_{E,lm}$ reads
\[
\begin{split}
\langle a_{E,l_1 m_1} a_{E,l_2 m_2} a_{E,l_3 m_3}\rangle_{\rm dyn}^{\rm (stan)}  =
&\ -  i^{l_1+l_2+l_3} \nico(l_1,2) \nico(l_2,2) \nico(l_3,2)  \frac{4}{3 H_0^2\Omega_m}
\\
& \times\int_0^{\chiS} \! d  \chi_1 d  \chi_2 d  \chi_3
W (\chiS, \chi_1)  W (\chiS, \chi_2)  W (\chiS, \chi_3) g(\chi_3) a(\chi_3)
\\
&\times
\int \frac{d^3\vec k_1}{2 \pi^2} \frac{d^3\vec k_2}{2 \pi^2}  \frac{d^3 \vec k_3}{2 \pi^2}
j_{l_1}(k_1 \chi_1) j_{l_2}(k_2 \chi_2) j_{l_3}(k_3 \chi_3)
Y^*_{l_1m_1}(\hat k_1) Y^*_{l_2m_2}(\hat k_2) Y^*_{l_3m_3}(\hat k_3)
\\
& \times T^2(k_1) T^2(k_2) \frac{k_1^2 k_2^2}{k^2_3}
F_{2,N}({\vec k_1},{\vec k_2},\chi_3)
P( k_1) P( k_2)
(2 \pi)^3 \delta_D( \vec k_1 + \vec k_2 +\vec k_3)
\\
+&\ \hbox{5 perms}\;. \label{stand_dyn1}
\end{split}
\]
Now we can write the Dirac delta function in the last line of this expression as
\[
\begin{split}
\label{delta_exp}
\delta_D(\vec k_1+\vec k_2 +\vec k_3)=
&\ \frac{1}{(2\pi)^3}\int d^3 \vec y\ e^{i(\vec k_1 +\vec k_2 + \vec k_3)\cdot\vec y}
=8\sum_{l'_i m'_i}i^{l'_1+l'_2+l'_3}(-1)^{l'_1+l'_2+l'_3}
{\cal G}_{l_1' l_2' l_3'}^{m_1' m_2' m_3'}
\\
& \times Y_{l'_1m'_1}(\hat k_1)Y_{l'_2m'_2}(\hat k_2)Y_{l'_3m'_3}(\hat k_3)
\int_0^{\infty}d \chi \chi^2 j_{l'_1}(k_1\chi)j_{l'_2}(k_2\chi)j_{l'_3}(k_3\chi)\; ,
\end{split}
\]
where we have used $\vec y \equiv \hat n \chi$.
Integrating eq.~\eqref{stand_dyn1}  over the angular directions $d^2 \hat k_i$ using the above equation and the orthogonality relations of the spherical harmonics, we find for the bispectrum
\[
\label{bdynnewt}
\begin{split}
[b^{\rm (stan)}_{\rm dyn}]_{l_1 l_2 l_3}  = \
& \nico(l_1,2) \nico(l_2,2) \nico(l_3,2)\frac{2}{3\Omega_m H_0^2}
 \int_0^{\chiS}  d\chi_1 d\chi_2 d\chi_3
 W(\chiS,\chi_1)W(\chiS,\chi_2) W(\chiS,\chi_3)g(\chi_3)a(\chi_3)
\\
& \times
\int \frac{2k_1^2dk_1}{\pi} \frac{2k_2^2dk_2}{\pi}
\frac{2k_3^2dk_3}{\pi}  \frac{k_1^2k_2^2}{k^2_3}
T^2(k_1)T^2(k_2) F_{2,N}(k_1,k_2,k_3;\chi_3) P(k_1) P(k_2)
\\
&\times \int_0^{\infty} d \chi \;\chi^2
j_{l_1}( k_1 \chi_1) j_{l_1}( k_1 \chi)
j_{l_2}( k_2 \chi_2) j_{l_2}( k_2 \chi)
j_{l_3}( k_3 \chi_3) j_{l_3}( k_3 \chi)
\\
+&\ \hbox{5 perms} \; ,
\end{split}
\]
where the kernel $F_{2,N}$ is a function of the amplitudes of the three momenta $k_1, k_2, k_3$, as easily shown using $\vec k_1 \cdot \vec k_2 = (k_3^2 - k_1^2- k_2^2)/2$ in eq.~\eqref{F2}.

We now move to the dynamical corrections, eq.~\eqref{gamma_corr_dyn}. For the rest of this section we will assume matter dominance, so that the growth suppression factor is $g=1$ and the window function reduces to $W(\chi,\chi') = (\chi - \chi')/(\chi \chi')$.  The first line of eq.~\eqref{gamma_corr_dyn} yields a real lensing potential.  For the scalar relativistic contribution coming from the first two terms in eq.~\eqref{gamma_corr_dyn}, i.e.~$(\phir+\psir)/2$, the computation is analogous to the standard scalar part above, the only difference being the kernel. The details of the calculation and the result can be found in appendix~\ref{app:dyn}.

The computation of the vector and tensor contributions is more involved since their kernels depend not only on the amplitude of $\vec k_1$ and $\vec k_2$ but also on their directions. We present here the computation of the vector modes. 
The computation of the tensor modes, which is similar, can be found in appendix~\ref{app:dyn}. 
The vector modes contain two types of terms. The first one comes from the first line of eq.~\eqref{gamma_corr_dyn}. Using eq.~\eqref{omega_four} and after some manipulations it can be written, in Fourier space, as
\[
-\frac12 \omega_r (k_3)  = -\frac{i 2 a^{1/2}}{3 H_0 }
\frac{T(k_1) T(k_2)}{k_3^2}\Phi_{\vec k_1} \Phi_{\vec k_2}
\left[\frac{k_3^2(k_1^2+k_2^2)-(k_1^2 -k_2^2)^2}{2k_3^2} (\vec k_3\cdot\hat n)
 - k_2^2  (\vec k_1 \cdot \hat n )- k_1^2 (\vec k_2 \cdot \hat n) \right]\;. \label{vecrr}
\]
The computation of the term proportional to $\vec k_3\cdot\hat n$ is analogous to the computation of the Doppler term in the redshift correction.
Using that
\[
\label{exp_k}
i \vec k_3 \cdot \hat n\, e^{i \vec k_3 \cdot \hat n \chi}=  \partial_{\chi}e^{i \vec k_3 \cdot \hat n \chi}\;,
\]
 the time derivative of Fourier modes translates into a time derivative of spherical Bessel functions in the expansion of eq.~\eqref{delta_exp}.  The rest of the calculation follows the one for the scalar modes.

The same trick cannot be used for the second and third terms proportional to $\vec k_i \cdot \hat n, \ i=1,2$, since it would require to split the exponential as $e^{i \vec k_3 \cdot \vec x}=e^{i \vec k_1\cdot \vec x}e^{i \vec k_2\cdot \vec x}$. This would introduce additional infinite summations over $l'$ and $m'$ in the expansion of the exponential into spherical harmonics. The alternative solution is to expand directly the product $\vec k_i \cdot \hat n$ into spherical harmonics using
\[
\label{exp_qi}
\hat k_i \cdot \hat n=\frac{4\pi}{3}\sum_M Y^*_{1M}(\hat k_i)Y_{1M}(\hat n)\; .
\]
With this expansion we find that the $a_{E,lm}$ coefficient from the second term in the brackets of eq.~(\ref{vecrr}) reads
\[
\begin{split}
[a_{\rm dyn}^{(\rm corr)}]_{E,l_3 m_3 }=
&-\frac{32i}{9H_0}\nico(l,2)
\int_0^{\chiS}d\chi_3 W(\chiS,\chi_3)a^{1/2}(\chi_3)
\int \frac{k_3^2 dk_3}{2\pi^2}\frac{d^3 \vec k_1 d^3 \vec k_2 }{2\pi^2}
T(k_1) T(k_2)\Phi_{\vec k_1} \Phi_{\vec k_2}
\\
&\times \frac{k_1k_2^2}{k_3^2}
\sum_{l'_im'_i}
\sum_{M=-1,0,1}(-1)^{m_3+m'_1+m'_2+m_3'+l_3'}\; i^{l'_1+l'_2} \;
{\cal G}^{l'_1, l'_2, l_3' }_ {-m'_1, -m'_2,-m_3'}
{\cal G}^{l,l', 1}_{-m_3, m_3', M}
\\
& \times Y^*_{1M}(\hat k_1)Y_{l'_1m'_1}(-\hat k_1)Y_{l'_2m'_2}(-\hat k_2)
\int_0^{\infty} d \chi \chi^2
j_{l'_1}(k_1\chi)j_{l'_2}( k_2 \chi)j_{l'_3}(k_3 \chi)j_{l_3'}(k_3\chi_3) \; .
\end{split}
\]
When computing the 3-point function, one ends up with three spherical harmonics to integrate over $\hat k_1$, which give rise to another Gaunt integral.
The summation over three Gaunt integrals can then be simplified by using that
\begin{eqnarray}
\nonumber
&&\sum_{m'_1 m_3' M}(-1)^{m'_1+m_3'+M} \Gaunt {l'_1,} {l_2,} {l'}{-m'_1,}   {-m_2,}   {-m_3'}
\Gaunt {l_3,}   {l_3',} {1}  {-m_3,} {m_3',}  {M}   \Gaunt {l_1,}  {l'_1,}  {1} {-m_1,}  {m'_1,} {-M}
\\
\label{sumgaunt}
&&\qquad\qquad\qquad
=(-1)^{l'_1+l_3'+1}\sqrt{\frac{(2l_1+1)(2l_2+1)(2l_3+1)}{4\pi}}\frac{3(2l'_1+1)(2l_3'+1)}{4\pi}
\begin{pmatrix}l_1 & l_2 & l\\m_1 & m_2 & m\end{pmatrix}
\\
\nonumber
&&\qquad\qquad\qquad
\phantom{=}\times
\begin{pmatrix}l_1 & l'_1 & 1\\0 & 0 & 0\end{pmatrix}
\begin{pmatrix}l_3' & l_3 & 1\\0 & 0 & 0\end{pmatrix}
\begin{pmatrix}l_3' & l'_1 & l_2\\0 & 0 & 0\end{pmatrix}
\begin{Bmatrix}l_1&l_2&l_3\\l_3' &1&l'_1\end{Bmatrix}\;,
\end{eqnarray}
where $\begin{Bmatrix}l_1&l_2&l_3\\l'_1 &l'_2&l'_3\end{Bmatrix}$ denotes the 6-$j$ Wigner matrix.
The properties of this matrix impose that the only terms in the summation over $l'_1$ and $l'_3$ that give non-zero contributions are those with $l'_1=l_1\pm 1$ and $l'_3=l_3\pm 1$. Using the same procedure for the third term in eq.~(\ref{vecrr}), we find that the bispectrum of the vector contribution $-\frac12 \omega_r $ reads
\[
\begin{split}
\label{bdyncorrveclong}
[b_{\rm dyn}^{(\rm corr)}]_{l_1 l_2 l_3}=
&\frac{2}{3 H_0}\nico(l_1,2)\nico(l_2,2)\nico(l_3,2)
\int_0^{\chiS}d\chi_1 d\chi_2 d\chi_3
W(\chiS,\chi_1)W(\chiS,\chi_2)W(\chiS,\chi_3)a^{1/2}(\chi_3)
\\
& \times \int \frac{2 k_1^2dk_1}{\pi} \frac{2 k_2^2dk_2}{\pi} \frac{2 k_3^2dk_3}{\pi} T^2(k_1)T^2(k_2)
P(k_1) P(k_2)
\\
&\times\Bigg\{ \frac{(k_1^2+k_2^2)k_3^2-(k_1^2-k_2^2)^2}{2k_3^4}
\\
&\qquad\qquad\times
\int_0^{\infty} d \chi \chi^2
j_{l_1}(k_1\chi_1)j_{l_1}( k_1 \chi)j_{l_2}(k_2\chi_2)j_{l_2}( k_2 \chi)
\partial_{\chi_3}j_{l_3}(k_3\chi_3)j_{l_3}( k_3 \chi)
\\
&\qquad+2\frac{k_1k_2^2}{k_3^2}
\sum_{\substack{l'_1=l_1\pm 1\\l'_3=l_3\pm1}}(-1)^{l_2}i^{l_1'-l_1+1}(2l_1'+1)(2l_3'+1)
\\
&\qquad\qquad\times
\begin{Bmatrix}l_1&l_2&l_3\\l'_3 &1&l'_1\end{Bmatrix}
\begin{pmatrix}l_1 & l'_1 & 1\\0 & 0 & 0\end{pmatrix}
\begin{pmatrix}l_3 & l'_3 & 1\\0 & 0 &0\end{pmatrix}
\begin{pmatrix}l'_1 & l_2 & l'_3 \\ 0 & 0 & 0\end{pmatrix}
\begin{pmatrix}l_1 & l_2 & l_3 \\ 0 & 0 & 0\end{pmatrix}^{-1}
\\
& \qquad\qquad\times\int_0^{\infty} d \chi \chi^2j_{l_1}(k_1\chi_1)j_{l'_1}( k_1 \chi)j_{l_2}(k_2\chi_2)j_{l_2}( k_2 \chi)
j_{l'_3}(k_3\chi_3)j_{l'_3}( k_3 \chi)\Bigg\}
\\
&+\ \hbox{5 perms}\;.
\end{split}
\]

The second vector term comes from the second line of eq.~\eqref{gamma_corr_dyn} and reads
\[
\label{vectr}
\begin{split}
-\frac12 \spart {}_1\omega (\hat n \chi)   =
&  -\frac{i 2 a^{1/2}}{3 H_0}
\int \frac{d^3\vec k_3}{(2\pi)^3}\frac{d^3\vec k_1d^3\vec k_2}{(2\pi)^3}
\delta(\vec k_3-\vec k_1-\vec k_2)
\frac{T(k_1) T(k_2)}{k_3^2}\Phi_{\vec k_1} \Phi_{\vec k_2}
\\
&\times \spart \Bigg\{e^{i\vec k_3\cdot \hat n \chi}
\left[\frac{k_3^2(k_1^2+k_2^2)-(k_1^2-k_2^2)^2}{2k_3^2}(\vec k_3\cdot \hat e_+)
- k_1^2 (\vec k_2 \cdot \hat e_+) - k_2^2 (\vec k_1 \cdot \hat e_+ )
\right]\Bigg\}
\;.
\end{split}
\]
Here the new terms are those proportional to $\hat k_3 \cdot \hat e_+$. To compute these terms we rewrite them as 
$\hat k_3 \cdot \hat e_+=-\spart (\hat k_3 \cdot \hat n)$ and then we use the expansion in eq.~(\ref{exp_k}). The operator $\spart$ acts on the spherical harmonics giving a spin-1 spherical harmonic ${}_1 Y_{lm}(\vec n)$, that results in a Gaunt integral with spin 1. We also have terms proportional to $ \hat k_i\cdot \hat e_+,\; i=1,2$. Again, we rewrite these terms as $\hat k_i \cdot \hat e_+=-\spart (\hat k_i \cdot \hat n)$ and then we use the expansion in eq.~(\ref{exp_qi}). With this, we find for the bispectrum of the vector term $-\frac12 \spart {}_1\omega $,
\[
\begin{split}
\label{bdyncorrvectr}
[b_{\rm dyn}^{(\rm corr)}]_{l_1 l_2 l_3}=
&\frac{-2}{3 H_0}\nico(l_1,2)\nico(l_2,2)\nico(l_3,2)
\int_0^{\chiS}d\chi_1 d\chi_2 d\chi_3
W(\chi_1,\chiS)W(\chi_2,\chiS)\frac{\sqrt{a(\chi_3)}}{\chi_3}
\\
& \times \int
\frac{2 k_1^2dk_1}{\pi}
\frac{2 k_2^2dk_2}{\pi}
\frac{2 k_3^2dk_3}{\pi}
T^2(k_1)T^2(k_2) P(k_1) P(k_2)
\\
&\times
\Bigg\{ \frac{(k_1^2+k_2^2)k_3^2-(k_1^2-k_2^2)^2}{2k_3^4}
\\
&\qquad\quad\times
\int_0^{\infty} d \chi \chi^2
j_{l_1}(k_1\chi_1)j_{l_1}( k_1 \chi)
j_{l_2}(k_2\chi_2)j_{l_2}( k_2 \chi)
j_{l_3}(k_3\chi_3)j_{l_3}( k_3 \chi)
\\
&\phantom{\times\Bigg\{}
+2\frac{k_1k_2^2}{k_3^2}
\sum_{\substack{l'_1=l_1\pm 1\\l'_3=l_3\pm1}}
(-1)^{l_2}i^{l_1'-l_1+1}
\frac{(2l_1'+1)(2l_3'+1)}{\nico(l_3,2)^2}
\begin{Bmatrix}l_1&l_3&l_2\\l'_3 &l'_1&1\end{Bmatrix}
\begin{pmatrix}l_1 & l_2 & l_3 \\ 0 & 0 & 0\end{pmatrix}^{-1}
\\
&\qquad\quad\times \Bigg[4l'_3(l'_3+1)
\begin{pmatrix}l_1 & l'_1 & 1 \\ 0 & 0 & 0\end{pmatrix}
\begin{pmatrix}l_3 & l'_3 & 1 \\ 0 & 0 & 0\end{pmatrix}
\begin{pmatrix}l'_1 & l_2 & l'_3 \\ 0 & 0 & 0\end{pmatrix}
\\
&\qquad\qquad\quad+ \sqrt{2l'_3(l'_3+1)}(l^{'2}_3+l'_3+2)
\begin{pmatrix}l_1 & l'_1 & 1 \\ 0 & 0 & 0\end{pmatrix}
\begin{pmatrix}l_3 & l'_3 & 1 \\ 0 & 1 & -1\end{pmatrix}
\begin{pmatrix}l'_1 & l_2 & l'_3 \\ 0 & 0 & 0\end{pmatrix}  \Bigg]
\\
&\qquad\quad\times
\int_0^{\infty} d \chi \chi^2
j_{l_1}(k_1\chi_1)j_{l'_1}( k_1 \chi)
j_{l_2}(k_2\chi_2)j_{l_2}( k_2 \chi)
j_{l'_3}(k_3\chi_3)j_{l'_3}( k_3 \chi)
\Bigg\}\\
+&\ \hbox{5 perms}.
\end{split}
\]

The computation of the bispectrum from the tensor contribution is analogous to the one above and is reported in appendix~\ref{app:dyn}.

\clearpage
\section{Results}
\label{sec:results}

\begin{figure}[h]
\begin{center}
{\includegraphics[width=0.4\textwidth]{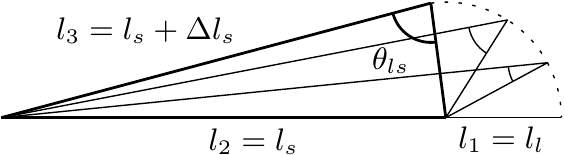}}
\caption{{Possible choice of $l_1$, $l_2$ and $l_3$ in the squeezed limit. Because of the triangle inequality enforced by the Gaunt integral, we let $\Delta \ls$ vary from $0$ to $\lL$.}}
\label{fig:triangle}
\end{center}
\end{figure}

We now evaluate the weighted bispectrum from the geometrical and the dynamical terms in a $\Lambda$CDM universe. Since we are mainly interested in the impact of the full-sky corrections, which become relevant when one of the modes becomes very long (of the order of the Hubble radius), we compute the bispectrum in the squeezed limit, $l_1 \ll l_2 \simeq l_3$. We label the low multipole corresponding to the long wavelength by $\lL$, and the two large multipoles, corresponding to the short wavelengths, by $\ls$ and $\ls+\Delta \ls$ respectively (see fig.~\ref{fig:triangle}).

\begin{table}[h]
\begin{center}
\small{\begin{tabular}{|l|c|c|}
\hline
Standard dynamical: newtonian kernel & $\hb_{\rm dyn}^{\rm (stan)}$, eq.~\eqref{bdynnewt} & {\color{red}\bf red} \\ \hline
Standard geometrical: lens-lens, Born corr., reduced shear & $\hb_{\rm geom}^{\rm (stan)}$, eq.~\eqref{bgeomstand} & {\color{blue}\bf blue} \\ \hline
Redshift corrections: Doppler, Sachs-Wolfe, ISW & $\hb_{\rm geom}^{\rm (z)}$, eq.~\eqref{bgeomz} & {\color{magenta}\bf magenta} \\ \hline
Geometrical corrections & $\hb_{\rm geom}^{\rm (corr)}$, eq.~\eqref{bgeomcorr} & {\color{black}\bf black} \\ \hline
Dynamical corrections: scalar, vector and tensor & $\hb_{\rm dyn}^{\rm (corr)}$ eqs.~\eqref{bdyncorrscal}, \eqref{bdyncorrveclong},&  {\color{green}\bf green} \\
 &
\eqref{bdyncorrtenslong}, \eqref{bdyncorrvectr}, \eqref{bdyncorrtenstr} and \eqref{bdyncorrtenssource} &  \\ \hline
\end{tabular}}
\end{center}
\caption {Color coding the different contributions to the bispectrum  showed in fig.~\ref{fig:bz1}  \label{t:color}}
\end{table}

\begin{figure}[t]
\begin{center}
{\includegraphics[width=1\textwidth]{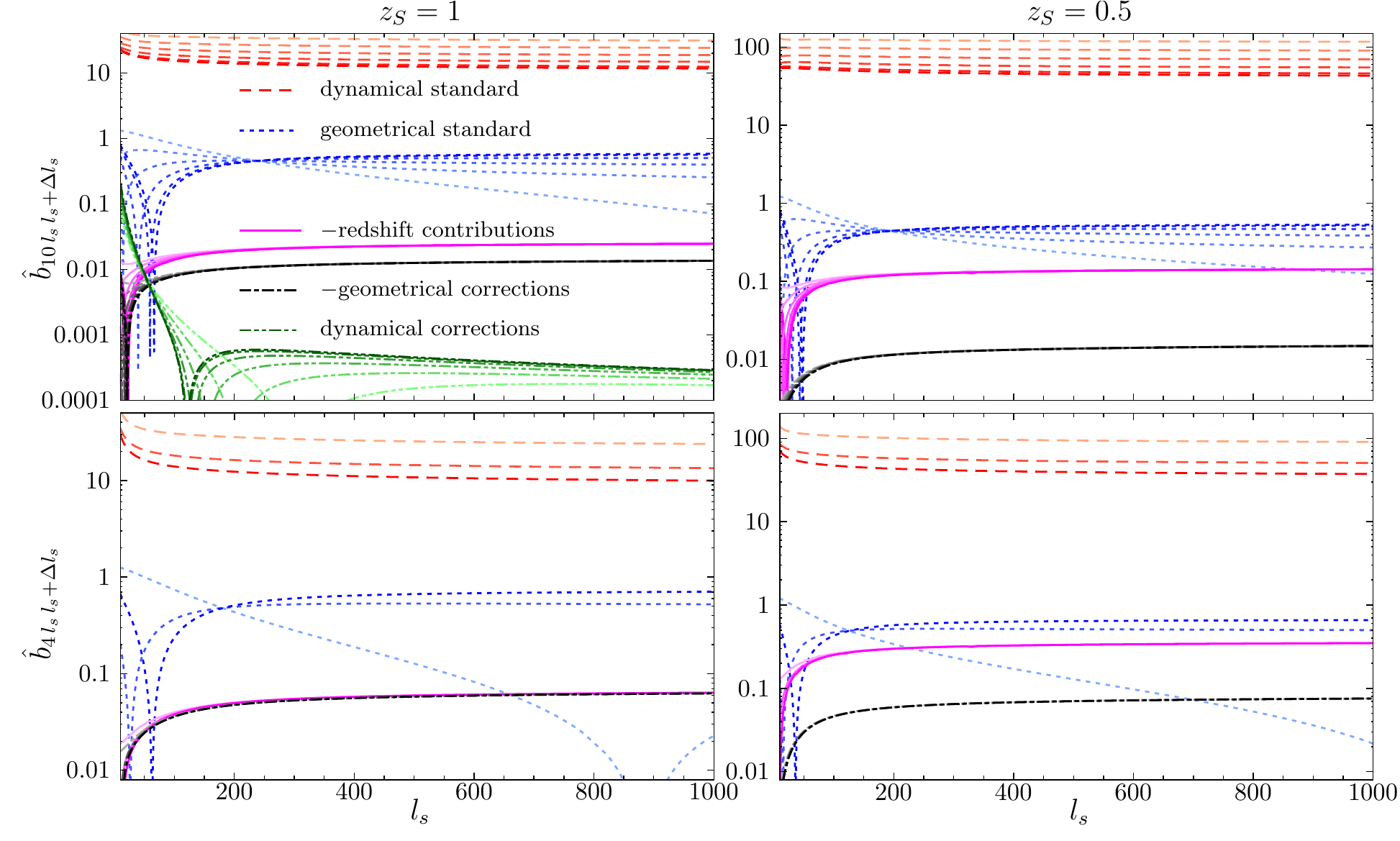}}
\caption{{The weighted bispectrum in the squeezed limit as a function of $\ls$, for two different values of the small multipole $\lL$ and two different source-redshifts, i.e.~$z_S=1$ (left-hand panels) and $z_S=0.5$ (right-hand panels). In the top panels we have plotted the weighted bispectrum
for $\lL=10$ and $ 0 \le  \Delta \ls \le 10$ while in the bottom panels we have plotted the case for $\lL=4$ and $  0 \le  \Delta \ls \le 4$.
The dynamical standard terms are in red, the geometrical standard terms in blue, the redshift contribution in magenta, the geometrical corrections in black and the dynamical corrections in green. For each term $\Delta \ls$ varies from 0 (darker line) to its maximum (lighter line) by steps of 2. The sign in the legend corresponds to the sign in the large-$\ls$ limit.}}
\label{fig:bz1}
\end{center}
\end{figure}
In fig.~\ref{fig:bz1} we plot the various contributions to the weighted bispectrum for two different values of the small multipole $\lL$ and two different values of the redshift of the sources $z_S$.
The amplitudes of the different contributions depend on  the redshift of the sources and on the value of  the small multipole $\lL$ and, as shown by this figure, there is a clear hierarchy between them.

\begin{figure}[hh]
\begin{center}
{\includegraphics[width=0.6\textwidth]{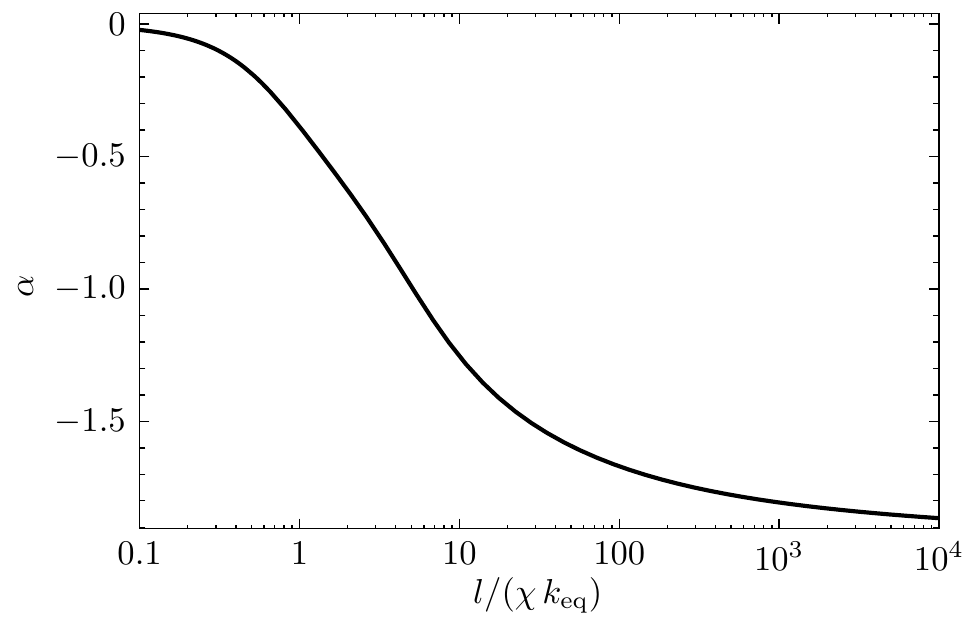}}
\caption{{The power $\alpha \equiv d \ln T /d \ln k$ as a function of $k/\keq = l/(\chi\, \keq)$. For the cosmology chosen in this paper ($h=0.65$ and $\Omega_m=0.3$) $\keq = 42.7 \; H_0$ and $\chi(z=1) = 0.77 \; H_0^{-1}$ so that for $l \sim 1000$, $\alpha \sim -1.5$.}}
\label{fig:alpha}
\end{center}
\end{figure}

In the following we are going to discuss their amplitude and their shape, including their redshift dependence. Even though the bispectra shown in fig.~\ref{fig:bz1} were computed using the second-order Limber approximation, for the following analysis we will make use of the first-order Limber approximation. Moreover, we will assume that  the matter transfer function $T(k)$ can be approximated by a power law,
\[
T(k=\nu/\chi) \propto \left(\frac{l}{\chi\ \keq}\right)^\alpha\; ,
\]
where $\keq$ is the equality scale. Clearly, as shown in fig.~\ref{fig:alpha} the power $\alpha$ depends on the scales that we consider.  At very small $l$'s, $\alpha \to 0$ for a scale invariant spectrum.
Using this ansatz, we can employ eq.~\eqref{Cl_limber} to derive how the angular power spectrum $C_l$ scales with $l$ and with the 
distance of the observer to the source, $\chiS$. We find
\[
 C_l \sim l \left( \frac{l}{\chi_S \keq} \right)^{2 \alpha} (1+ {\cal O} (l^{-2})) \;,   \label{LimberedCl}
\]
The Limber approximated formulae for the bispectrum can be found in appendix~\ref{app:limber}.
For a summary of the color coding in the figures and of the equations giving the expression of the various contributions see table~\ref{t:color}.

\subsection{Geometrical correction}

We will start by describing the geometrical relativistic correction, eq.~\eqref{bgeomcorr}, black line in fig.~\ref{fig:bz1}. This contains two terms that we study separately but will give identical behavior. The first term contains a total derivative $\spart^2\big(\Psi(\chi,\vec x)\Psi(\chi', \vec x')\big)$ and its Limber approximation is given by the first five lines of eq.~\eqref{bgeomcorrlimb1}. Using this equation we can find the scaling behavior of this term,
\[
b_{l_1 l_2 l_3} \sim \left(\frac{l_1}{\chi_S \keq} \right)^{2 \alpha_1} l_1^{-1} \left(\frac{l_2}{\chi_S \keq} \right)^{2 \alpha_2} l_2^{ -1} l_3^{2}   + 5 \ \text{perms} \;.
\]
Because the term above is a total gradient there is no modulation dependence on the angle between the short and long mode.
We can compute the weighted bispectrum in the squeezed limit by using the expression for the $C_l$ in eq.~\eqref{LimberedCl}. 
Choosing  $l_1=l_l$  gives
\[
\hat b_{\lL \ls \ls+\Delta \ls} \simeq \frac{b_{\lL \ls \ls+\Delta \ls}}{2 C_{\lL} C_{\ls}} \sim \lL^{-2}   \;\big( 1 + {\cal O} ( \ls^{-2})\big)\;. \label{scaling_gc}
\]
Thus,  the weighted bispectrum for  this term does not depend on the short mode $\ls$ and gives rise to a plateau at large $\ls$, as shown in fig.~\ref{fig:bz1}. Furthermore, as it does not depend on the angle between the short and long mode, it does not depend on $\Delta \ls$. However, its dependence on $\lL$ can be clearly seen by comparing the upper and lower panel in fig.~\ref{fig:bz1}. As expected from eq.~\eqref{scaling_gc}, this contribution is large from smaller $\lL$.

\begin{figure}[t]
\begin{center}
{\includegraphics[width=0.60\textwidth]{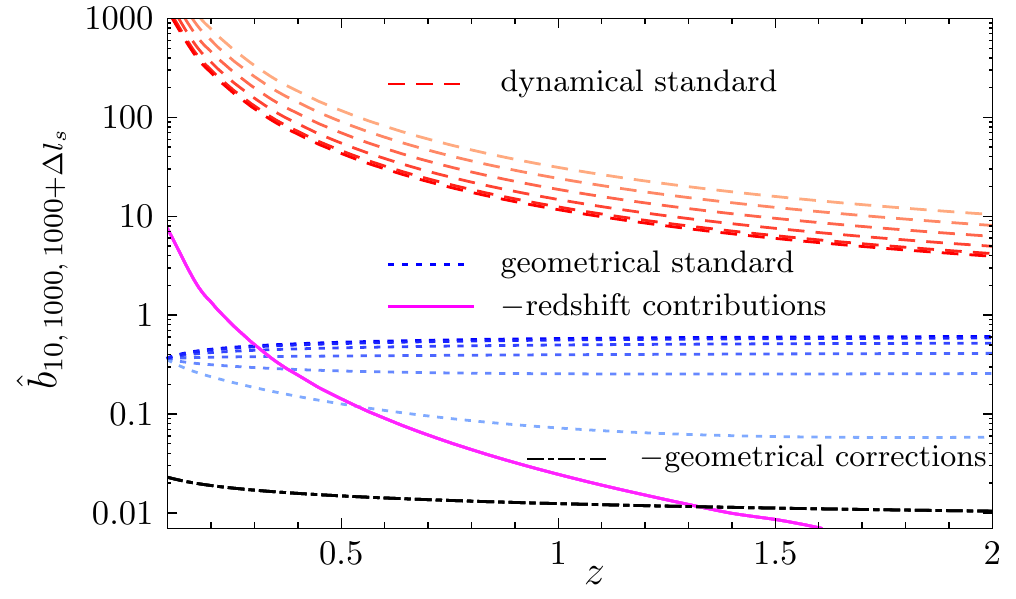}}
\caption{{The weighted bispectrum in the configuration $\lL=10$, $\ls=1000$ and $0 \le \Delta \ls \le 10$, plotted as a function of the redshift of the source $z_S$. The color coding is the same as in fig.~\ref{fig:bz1}}}
\label{fig:bzp0}
\end{center}
\end{figure}

Note that the dependence on $\chiS$ drops out in the weighted bispectrum so that the latter depends very mildly on the redshift of the sources $z_S$, as can be seen by comparing the left- and right-hand panels in fig.~\ref{fig:bz1}.
In order to better visualize the redshift dependence, in fig.~\ref{fig:bzp0} we  have plotted the weighted bispectrum for a particular configuration ($\lL=10$, $\ls=1000$ and $0 \le \Delta \ls \le 10$) as a function of $z_S$.

Let us discuss now the second term, containing $\Psi(\chi,\vec x)\spart^2\Psi(\chi',\vec x')$, i.e.~a coupling of a spin-0 with a spin-2 field. Its Limber approximation is given by the last five lines of eq.~\eqref{bgeomcorrlimb1}. By applying the same scaling argument as above we obtain
\[
b_{l_1 l_2 l_3} \sim \left(\frac{l_1}{\chi_S \keq} \right)^{2 \alpha_1} l_1^{-1} \left(\frac{l_2}{\chi_S \keq} \right)^{2 \alpha_2} l_2  \cos ( 2 \theta_{23})  + 5 \ \text{perms} \;, \label{gc_1}
\]
where $\theta_{23}$ is the angle between $l_2$ and $l_3$ {\em inside} the triangle, and we have used the general expression
\[
\label{cos}
 \begin{pmatrix}l_1 & l_2 & l_3\\ 0 & -s & s\end{pmatrix} \begin{pmatrix}l_1 & l_2 & l_3\\0 & 0 & 0\end{pmatrix} ^{-1} \approx
(-1)^s \cos{(s\cdot \theta_{23})}~,
\]
valid for large $l$'s.
If in the squeezed limit we take $l_1 = \lL$ and $l_2 \simeq l_3 = \ls$, the dominant term in the permutation is the first on the right-hand side of eq.~\eqref{gc_1} or the equivalent term with $2$ and $3$ exchanged. These are proportional to $\cos ( 2 \theta_{23})$ -- where $\theta_{23}$ is the angle between the two short wavemodes -- which in the squeezed limit goes to $1$. Thus, also for this term we expect  no modulation on the angle between the short and long mode and
for the weighted bispectrum we find the same scaling as for the first term, eq.~\eqref{scaling_gc}. Again, the dependence on $\chiS$ drops out in  the weighted bispectrum, whence it does not depend on the redshift.

\subsection{Redshift correction}

Let us discuss now the redshift correction, eq.~\eqref{bgeomz}, in magenta in fig.~\ref{fig:bz1}.
This contribution is of the form $\Psi(\chi,\vec x)\spart^2\Psi(\chi',\vec x')($\footnote{The Doppler term contains an additional gradient of $\Psi$, but this has no impact on the Wigner symbols and consequently on the $\Delta \ls$ dependence of this contribution.}) and the same arguments as for the second term above can be applied (the first-order Limber approximation for the redshift correction is given in eq.~\eqref{bzcorrlimb1}). Hence, the redshift
correction is also constant in $\ls$ and is independent of $\Delta \ls$, as shown in fig.~\ref{fig:bz1} (magenta lines).
However, the redshift dependence of this contribution is more complex than the one of the geometrical correction and we need to study it in more details.

Indeed, as shown in eq.~\eqref{geom_z}, the redshift correction is made of three contributions: a Sachs-Wolfe, an integrated Sachs-Wolfe and a Doppler term.
Using eq.~\eqref{bgeomz} we can infer the scaling with the source distance from the observer. In the squeezed limit the first two contributions scale as
\[
\hat b_{\lL \ls \ls+\Delta \ls} \sim
\lL^{-2} \;  \frac{\etaS}{\chiS}   \;( 1 + {\cal O} ( \ls^{-2})) \;,
\]
where  $\etaS$ is the conformal time at the source. The extra factor ${\etaS}/{\chiS}$ with respect to the geometrical correction comes from the term in front of the first square bracket of eq.~\eqref{bgeomz}. For the Doppler term one finds
\[
\hat b_{\lL \ls \ls+\Delta \ls} \sim
\lL^{-2} \; \left( \frac{\etaS}{\chiS} \right)^2  \;( 1 + {\cal O} ( \ls^{-2}))  \;.
\]
Thus, for sources at very high redshifts, i.e.~such that $\etaS \lesssim  \chiS $,  the redshift correction is typically dominated by the Sachs-Wolfe and integrated Sachs-Wolfe term, and is subdominant with respect to the geometrical correction. For sources at low redshifts $\etaS \gtrsim  \chiS $ and the Doppler term dominates over the Sachs-Wolfe and integrated Sachs-Wolfe terms, and the redshift correction is larger than the geometrical correction. The redshift of transition is around $z_S \sim 1.2$ but its exact value depends on the particular configuration. This is shown by  comparing the left- and right-hand panels in fig.~\ref{fig:bz1} and more directly in fig.~\ref{fig:bzp0}, where for each contribution we plot a particular configuration as a function of the redshift $z_S$.

A remark is in order here. Naively one would think that the contribution coming from the Doppler term vanishes. Indeed, in the small angle approximation the correlation between the Doppler term and the first-order shear is taken at the same time $\chi=\chi_S$ and vanishes for two reasons:
the first-order shear  vanishes at the source because of the window function; the Doppler term is longitudinal to the line of sight while the shear is transverse, so that being orthogonal their correlation at the same time vanishes. 
However, the correlation of the lens with the source does not vanish immediately for $\chi < \chiS$, but for a given mode $l$ it remains large as long as $\chiS -\chi \lesssim \chi /l$, after which it decays exponentially (see appendix of \cite{Boubekeur:2008kn}). In other words, the correlation between the source and the lens decays when the distance from the source is of the order of the typical wavelength. This also implies that the redshift correction is dominated by the configuration in which the Doppler term and one of the first-order shear terms in eq.~\eqref{bgeomz} is in the long mode.

\subsection{Standard geometrical contribution}

We now study the standard geometrical  contribution, given in eq.~\eqref{bgeomstand} and  shown in blue in fig.~\ref{fig:bz1}. This contribution is  given by the collection of the lens-lens coupling, the Born correction, and the non-linear correction induced by going from the shear to the reduced shear. All these terms contain four transverse derivatives of the Weyl potential and we expect that they dominate over the two contributions discussed above.

\begin{figure}[!h]
\begin{center}
{\includegraphics[width=1\textwidth]{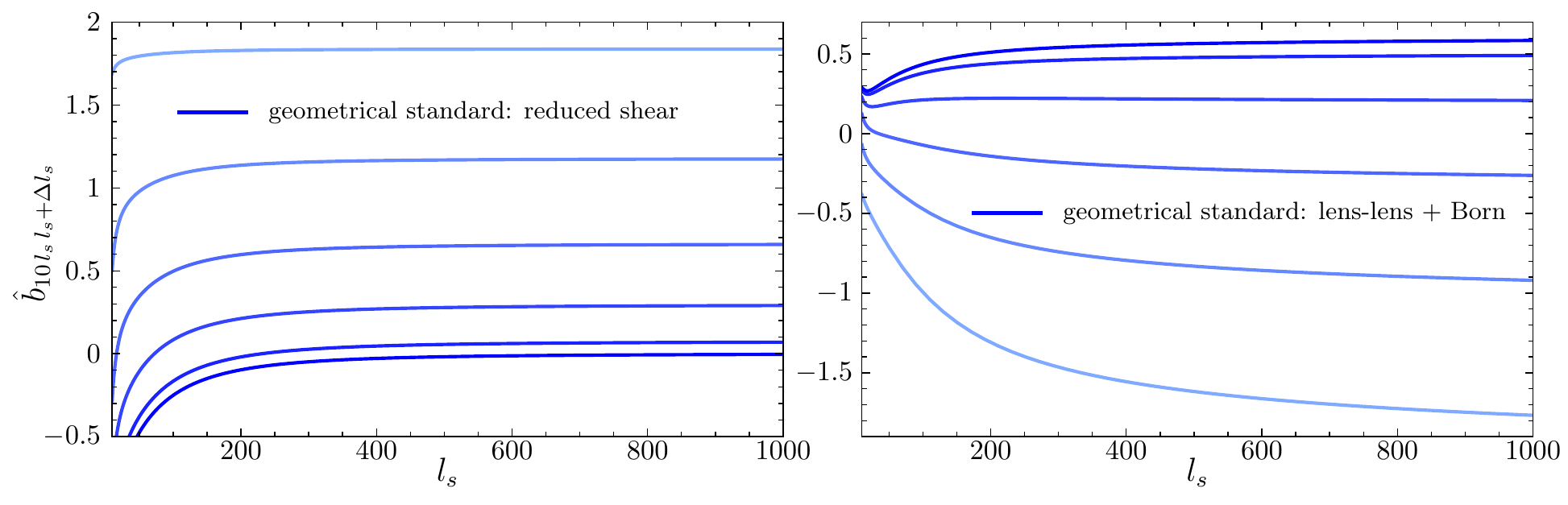}}
\caption{{The weighted bispectrum from the reduced shear contribution (left panel) and from the lens-lens coupling and the correction to the Born approximation (right panel) at $z_S=1$, in the squeezed limit: $\lL=10$ and $ 0 \le \Delta \ls  \le 10$. $\Delta \ls$ varies from $0$ (dark blue) to $10$ (light blue) by steps of 2.}}
\label{fig:bisp6}
\end{center}
\end{figure}

Let us start discussing the correction induced by the reduced shear. This term is proportional to  $\spart^2\Psi(\chi,\vec x)\spartb\spart\Psi(\chi',\vec x')$ (see e.g.~last line of eq.~\eqref{geom_stand}) and hence describes the coupling between a spin-2  and a spin-0 field.
Using the first-order Limber approximation given in the last two lines of eq.~\eqref{bgeomstandlimb1},  the bispectrum scales as
\[
b_{l_1 l_2 l_3} \sim  \left( \frac{l_1}{\chiS \keq} \right)^{2 \alpha_1} l_1 \left( \frac{l_2}{\chiS \keq} \right)^{2 \alpha_2} l_2 \cos(2 \theta_{23})  + \mbox{5 perms}~,
\]
where we have employed eq.~\eqref{cos} to rewrite the Wigner symbols in terms of the cosine.
If we chose $l_1 = \lL$, the four dominant configurations in the squeezed limit are those containing $l_1$, i.e., summing up all of them,
\[
b_{l_1 l_2 l_3} \sim  \left( \frac{l_1}{\chiS \keq} \right)^{2 \alpha_1} l_1 \left( \frac{l_2}{\chiS \keq} \right)^{2 \alpha_2} l_2 [ \cos(2 \theta_{13}) + \cos(2 \theta_{12}) + 2 \cos(2 \theta_{23}) ]    \;,
\]
where we have used $l_2 \simeq l_3$. In the squeezed limit $\theta_{23} \to 0$ and $\theta_{12} \to \pi - \theta_{13}$, so that we can write the scaling of the weighted bispectrum as
\[
\hat b_{\lL \ls \ls+\Delta \ls} \sim (1+ \cos (2 \thetals))   \; (1+ {\cal O}(\ls^{-2})) \;.
\]
Thus, this contribution is $\ls$-independent for large $\ls$'s but its shape depends on the angle between the long and the short mode. For $\Delta \ls \ll \ls$, $\cos \thetals \simeq \Delta \ls / \lL$ and $1+ \cos ( 2 \thetals) \simeq 2 (\Delta \ls / \lL)^2$. This can be seen in detail in fig.~\ref{fig:bisp6} (left panel) where we have plotted this contribution separately from the other geometrical standard contributions. One can check that in the large-$\ls$ limit the weighted bispectrum vanishes for $\Delta \ls=0$, and becomes positive when $\Delta \ls$ goes to $\lL$. 

We can now concentrate on the lens-lens coupling and the correction to Born approximation, which are described by the first two lines of eq.~\eqref{geom_stand}.
They contain three types of couplings: couplings between a spin-3  and a spin-$(-1)$ field, $\spart^3\Psi(\chi,\vec x)\spartb\Psi(\chi',\vec x')$, couplings between two spin-1 fields, $\spartb\spart^2\Psi(\chi,\vec x)\spart\Psi(\chi',\vec x')$, and couplings between a spin-2  and a spin-0 field, $\spart^2\Psi(\chi,\vec x)\spart\spartb\Psi(\chi',\vec x')$. These types of couplings generate three different Wigner symbols, written  in eq.~\eqref{bgeomstand}. In order to study the scaling of this contribution we can use the recurrence identities for the Wigner symbols to write
\[
\begin{split}
\begin{pmatrix}l_1 & l_2 & l_3\\3 & -1 & -2\end{pmatrix}=& - \sqrt{\frac{(l_3-2)(l_3+3)}{l_2(l_2+1)}}
\begin{pmatrix}l_1 & l_2 & l_3\\3 & 0 & -3\end{pmatrix}
-\sqrt{\frac{(l_1+3)(l_1-2)}{l_2(l_2+1)}}
\begin{pmatrix}l_1 & l_2 & l_3\\2 &0 & -2\end{pmatrix}\; ,
\\
\begin{pmatrix}l_1 & l_2 & l_3\\1 & 1 & -2\end{pmatrix}=&-\sqrt{\frac{(l_3+2)(l_3-1)}{l_2(l_2+1)}}
\begin{pmatrix}l_1 & l_2 & l_3\\1 & 0 & -1\end{pmatrix}
-\sqrt{\frac{(l_1-1)(l_1+2)}{l_2(l_2+1)}}
\begin{pmatrix}l_1 & l_2 & l_3\\2 &0 & -2\end{pmatrix}~,
\end{split}
\]
so that in the large $l$'s limit we can rewrite all the Wigner symbols in terms of cosines using eq.~\eqref{cos}.
Thus, inserting these decompositions into the first-order Limber approximation of the bispectrum given by the first seven lines of eq.~\eqref{bgeomstandlimb1}, the lens-lens coupling and the correction to the Born approximation scale as
\[
\begin{split}
b_{l_1 l_2 l_3} \sim
\left( \frac{l_1}{\chiS \keq} \right)^{2 \alpha_1} l_1
\left( \frac{l_2}{\chiS \keq} \right)^{2 \alpha_2} l_2
\left[
\frac{l_1 l_3}{l_2^2} \left( \cos(3 \theta_{13}) + \cos( \theta_{13}) \right) 
-  2 \frac{ l_1^2}{l_2^2}  \cos(2 \theta_{13})
+ \cos(2 \theta_{13})  + \cos(2 \theta_{23})
\right]
\\
+ 5 \ \text{perms}
\;.
\end{split}
\]
Taking the squeezed limit of this expression is subtle: at lowest order in $1/\ls$ (i.e.~at zeroth order in  the expansion of the cosines in terms of the small angle between the two short wavemodes) the expression above vanishes. Thus, we need to compute this expression at least at an order ${\cal O}(\ls^{-2})$ higher.  Taking into account  all the permutations one finds
\[
\hat b_{\lL \ls \ls+\Delta \ls} \sim (1- 6 \cos^2 \thetals)  \; (1+ {\cal O}(\ls^{-2})) \;.
\]
This contribution is shown in details in fig.~\ref{fig:bisp6} (right panel). Using that $1 - 6 \cos^2 \thetals  \simeq  1- 6 {\Delta \ls^2}/{\lL^2}$ one can check that it is positive for $\Delta \ls \le 4$.

Summing the reduced shear contribution to the lens-lens coupling and the Born correction, we find that the standard geometrical contribution gives rise to a plateau
as a function of $\ls$. The redshift dependence of the standard geometrical  contribution is similar to the  one of the geometrical correction.  However, its amplitude is much larger due to the absence of the $\lL^{-2}$ suppression present in the geometrical correction. Moreover, it depends on $\Delta \ls$, as shown in fig.~\ref{fig:bz1} (blue lines).
As shown in the lower-right panel of fig.~\ref{fig:bz1}, for  very small $\lL$ and low $z_S$, the redshift correction becomes of the same order as the standard geometric contribution.

\subsection{Dynamical contributions}

The largest contribution to the cosmic shear comes from what we have called the standard dynamical contribution. This is given in eq.~\eqref{bdynnewt} and  shown in red in fig.~\ref{fig:bz1}. This contribution is proportional to the total transverse Laplacian of a scalar, $\spart^2 \phi_N^{(2)}(\chi,\vec x)$ (see eq.~\eqref{dyn_stand}), but the scalar $\phi_N^{(2)}$ is not a Gaussian variable: it is related to the Gaussian primordial perturbation $\Phi$ by eqs.~\eqref{phinewt} and \eqref{F2}. We can compute the scaling of this  contribution as we did  above for the others. We obtain
\[
b_{l_1 l_2 l_3} \sim   \left( \frac{l_1}{\chiS \keq} \right)^{2 \alpha_1} l_1 \left( \frac{l_2}{\chiS \keq} \right)^{2 \alpha_2} l_2  \left(  \frac{\eta_0}{\chiS}\right)^2 F_{2,N} (\vec l_1, \vec l_2 ) + 5 \ \text{perms}\;,
\]
where, from eq.~\eqref{F2},
\[
F_{2,N} (\vec l_1, \vec l_2) \equiv \frac12 (1+\epsilon) + \frac12 \left( \frac{l_1}{l_2} + \frac{l_2}{l_1}\right) \cos \theta_{12} + \frac12 (1- \epsilon) \cos^2 \theta_{12} \;.
\]

If we take the squeezed limit by choosing $\lL = l_1$, there are four contributions that dominate,
\[
b_{l_1 l_2 l_3} \sim 
 \left( \frac{l_1}{\chiS \keq} \right)^{2 \alpha_1} l_1 \left( \frac{l_2}{\chiS \keq} \right)^{2 \alpha_2} l_2
\left(2F_{2,N} (\vec l_1,\vec l_2 ) +2  F_{2,N} (\vec l_1,\vec l_3 )\right)
\left(\frac{\eta_0}{\chiS}\right)^2\;,
\]
where we have used $l_2 \simeq l_3$. As in the squeezed limit $\cos \theta_{13} = - \cos \theta_{12}$, the terms proportional to $\cos \theta_{12}$  and $\cos \theta_{13}$ cancel and we finally obtain
\[
\hat b_{\lL \ls \ls+\Delta \ls} \sim 2 \left[ (1+\epsilon) + (1-\epsilon) \cos^2 \thetals \right]  \left(  \frac{\eta_0}{\chiS}\right)^2  \; (1+ {\cal O}(\ls^{-2})) \;.
\]
Thus, as can be checked in fig.~\ref{fig:bz1},  the weighted bispectrum of the standard dynamical contribution does not depend on $\ls$ but depends on the angle between the short and the long mode.  This contribution dominates over the standard geometrical contribution because it is enhanced by the term $({\eta_0}/{\chiS})^2$. This term decreases at higher redshift, as shown by comparing the left and right panels of fig.~\ref{fig:bz1} and more clearly by fig.~\ref{fig:bzp0}.

The dynamical correction from scalar, vector and tensor modes, see eqs.~\eqref{bdyncorrscal}, \eqref{bdyncorrveclong},
\eqref{bdyncorrtenslong}, \eqref{bdyncorrvectr}, \eqref{bdyncorrtenstr} and \eqref{bdyncorrtenssource}, is shown in green in fig.~\ref{fig:bz1}, left-top panel. For large $\ls$ it is at least one order of magnitude smaller than the geometrical correction. Only for $\ls \lesssim 100$  it becomes larger than the geometrical correction and the redshift contribution.
We will neglect it in the analysis of the following section.

\clearpage
\section{Comparison with primordial non-Gaussianity}
\label{sec:primordial}

\subsection{Local bispectrum}

Let us start by studying the effect that a primordial signal would induce on the lensing shear bispectrum.
The most common type of non-Gaussianity considered in the literature is the  local type, which is typically generated by multi-field models \cite{Lyth:2001nq,Bernardeau:2002jy,Dvali:2003ar,Vernizzi:2006ve}. For local non-Gaussianity, the first-order primordial potential $\Phi$ (which in matter dominance is related to the primordial curvature perturbation $\zeta$ by $\Phi= - (3/5) \zeta$) contains a primordial quadratic correction in position space \cite{1990PhRvD..42.3936S},
\[
\Phi(\vec x) = \Phi_G(\vec x) - \fNLloc \Phi_G^2(\vec x)\;, \label{local}
\]
where $\Phi_G$ is a primordial Gaussian potential whose power spectrum is given by eqs.~\eqref{spectrum_Phi} and \eqref{Pk}
\footnote{Note that we use the notation of~\cite{Sefusatti:2007ih} with a minus sign for $\fNL$. This is related to $\fNL$ used in CMB computation (which is defined in radiation domination) by 
$\fNL=10/9\fNL^{\rm{CMB}}$.}.
Using the above equation together with eqs.~\eqref{spectrum_Phi} and \eqref{Pk}, we can rewrite the 3-point correlation function of $\Phi$ in Fourier space as
\[
\langle \Phi_{\vec k_1} \Phi_{\vec k_2} \Phi_{\vec k_3} \rangle = (2\pi)^3 \delta(\vec k_1+ \vec k_2 + \vec k_3) B^{\rm loc} ( k_1, k_2 , k_3) \;, \label{3-point_local}
\]
where the local bispectrum is given by
\[
B^{\rm loc} ( k_1,  k_2 , k_3) = -2 \fNLloc A_\Phi^2 \left( \frac1{k_1^3 k_2^3} + \frac1{k_1^3 k_3^3} + \frac1{k_2^3 k_3^3} \right)\;. \label{Phi_bispectrum_loc}
\]
We can use eq.~\eqref{aE_lin} to compute the angular lensing bispectrum generated by local non-Gaussianity.
It is straightforward to show that this is given by
\[
\begin{split}
b_{l_1l_2l_3}^{\rm loc} = &\ -\nico(l_1,2) \nico(l_2,2) \nico(l_3,2)
\int_0^{\chiS} \! d\chi_1\, d\chi_2\, d\chi_3\,
W(\chiS,\chi_1)\,  W(\chiS,\chi_2)\,  W(\chiS,\chi_3)
\int_0^{\chiS} \! d\chi \; \chi^2 \\
&\times\int \frac{2 k_1^2 d k_1}{\pi} \frac{2 k_2^2 d k_2}{\pi}\frac{2 k_3^2 d k_3}{\pi} j_{l_1}(k_1 \chi) j_{l_2}(k_2 \chi) j_{l_3}(k_3 \chi)
j_{l_1}(k_1 \chi_1) j_{l_2}(k_2 \chi_2) j_{l_3}(k_3 \chi_3)   \\
& \times T(k_1)\,T(k_2)\,T(k_3)\,B^{\rm loc}(k_1, k_2 ,  k_3)\;. \label{local_bispectrum}
\end{split}
\]

Let us use the Limber approximation and the scaling arguments of the previous section to find the behavior of the local bispectrum as a function of the long and short modes. Employing again the relation $T(k) \propto l^\alpha$, one finds
\[
b_{l_1l_2l_3}^{\rm loc} \sim  \left(   \frac{l_1}{\chiS \keq} \right)^{ \alpha_1} l_1^{ -1} \left(   \frac{l_2}{\chiS \keq} \right)^{ \alpha_2} l_2^{ -1}  \left( \frac{l_3}{\chiS \keq} \right)^{\alpha_3} l_3^2  + 5 \ \text{perms}\;.
\]
In the squeezed limit we obtain, for the weighted  bispectrum,
\[
\hat b_{\lL \ls \ls + \Delta \ls}^{\rm loc} \sim  \lL^{-2 }  \left( \frac{\lL}{\chiS \keq} \right)^{-\alpha_l} (1 + {\cal O}(\ls^{-2} ))\;.
\]
Thus, on very large scales, where $\alpha_l \to 0$, the local bispectrum behaves as the geometrical correction.

\subsection{Contamination}

The estimator of local primordial non-Gaussianity is given by \cite{Komatsu:2003iq}
\[
{\cal E} = \frac{1}{N_{\rm loc}} \sum_{l_i m_i}
\begin{pmatrix}l_1 & l_2 & l_3\\m_1 & m_2 & m_3\end{pmatrix}
\frac{B_{l_1 l_2 l_3}^{\rm loc}}{\Delta_{l_1 l_2 l_3}C_{l_1} C_{l_2} C_{l_3}}
\left\langle a_{E,l_1 m_1} a_{E,l_2 m_2} a_{E,l_3 m_3}\right\rangle\;, \label{estimator}
\]
where $ \Delta_{l_1 l_2 l_3}$	is a combinatorial factor equal to 1 if the three $l$'s are different, to 2 if two of them are equal and to 6 if all of them are equal. If the local bispectrum $B_{l_1 l_2 l_3}^{\rm loc}$ is computed for $\fNLloc=1$, the normalization factor $N_{\rm loc}$ that makes the estimator unbiased ($\langle {\cal E} \rangle= \fNLloc$) is
\[
N_{\rm loc} = \sum_{l_1 l_2 l_3} \frac{(B_{l_1 l_2 l_3}^{\rm loc})^2}{\Delta_{l_1 l_2 l_3} C_{l_1} C_{l_2} C_{l_3}}\;. \label{N_unbiased}
\]
Thus, if this estimator is applied to an arbitrary signal $B^{X}_{l_1 l_2 l_3}$, this will contaminate the measurement by a value of $\fNLloc$ given by
\[
\fNLloc= \frac{1}{N_{\rm loc}} \sum_{l_1 l_2 l_3}  \frac{B_{l_1 l_2 l_3}^{\rm loc} B_{l_1 l_2 l_3}^{X} }{\Delta_{l_1 l_2 l_3} C_{l_1} C_{l_2} C_{l_3}} \;. \label{contamination}
\]

\begin{figure}[t]
\begin{center}
{\includegraphics[width=0.6\textwidth]{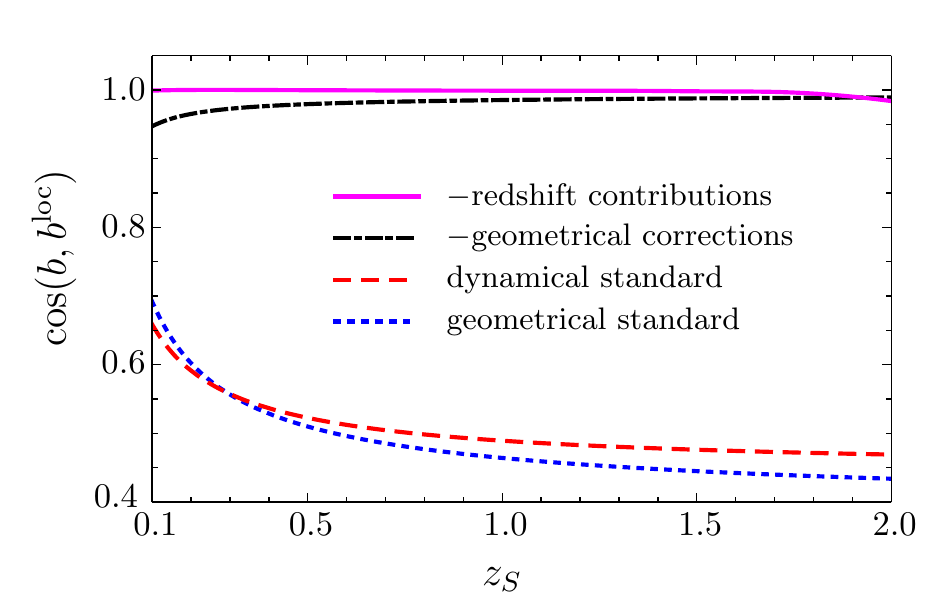}}
\caption{{Overlap between the local non-Gaussianity and the standard  dynamical and geometrical contributions and the geometrical and redshift corrections.}}
\label{fig:cos}
\end{center}
\end{figure}
Before estimating the contamination to the local signal from each of the contributions studied in the previous sections, let us use eq.~\eqref{estimator} to define a natural quantity that characterizes the ``superposition'' between two different contributions. We define the cosine between two bispectra as \cite{Babich:2004gb}
\[
\cos (Y , X) \equiv \frac{1}{N_{X}^{1/2} N_{Y}^{1/2} } \sum_{l_1 l_2 l_3} \frac{B_{l_1 l_2 l_3}^{X} B_{l_1 l_2 l_3}^{Y} }{\Delta_{l_1 l_2 l_3} C_{l_1} C_{l_2} C_{l_3}}\;,
\]
where the factors $N_{X,Y}$ are defined analogously to $N_{\rm loc}$ in eq.~\eqref{N_unbiased}. This quantity ranges from $-1$ to $1$ and its absolute value tells  us how much two signals are orthogonal, $\cos (Y, X)  \simeq 0$, or superposed, $ \cos (Y ,X) \simeq \pm 1$. The cosine between the local non-Gaussianity and the contributions to the lensing shear from second-order perturbations is plotted in fig.~\ref{fig:cos} as a function of the redshift. For the calculation, we have summed over all $l$'s (with even sum) from $2$ to $l_{\rm max} = 1000$. As expected, the geometrical relativistic corrections and the correction due to the redshift perturbation overlap very much with a local non-Gaussianity, contributing with a negative $f^{\rm loc}_{\rm NL}$. Indeed,  in the squeezed limit both these corrections have the same $l$ behavior and no dependence on the angle between the short and long mode. The overlap with the standard contributions is much milder. This is due to the fact that  in the equilateral limit its amplitude decays less rapidly than the one of the local shape and because part of the standard contributions depend on the angle between the short and long mode, and this part averages to zero when convolved with the local signal.

\begin{figure}[t]
\begin{center}
{\includegraphics[width=0.6\textwidth]{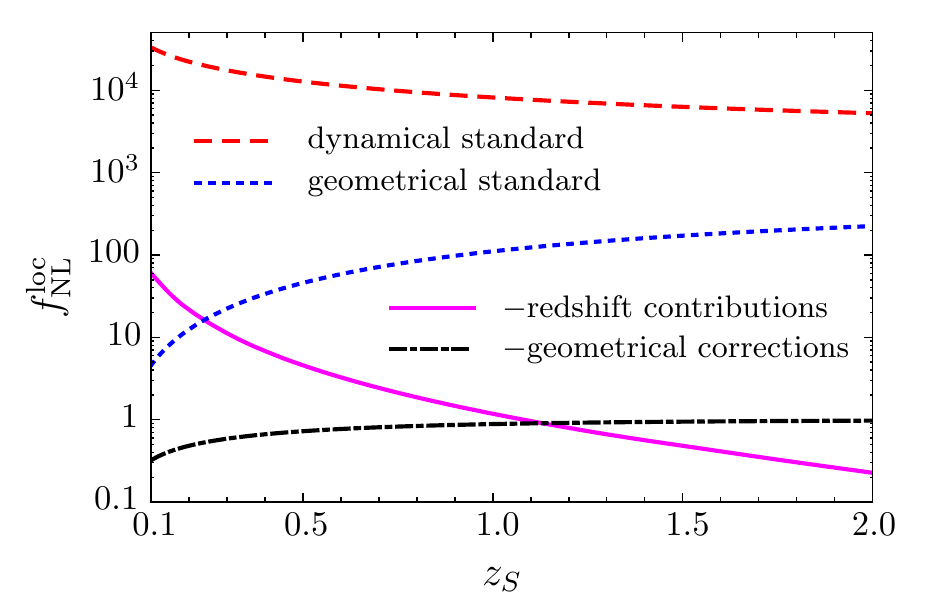}}
\caption{{The contamination to a primordial local non-Gaussian signal due to  the standard  dynamical and geometrical contributions and the geometrical and redshift corrections.}}
\label{fig:fNL}
\end{center}
\end{figure}
We can now use eq.~\eqref{contamination} to compute the non-Gaussian contamination to the local signal from each of these four contributions. The $\fNLloc$ that an experiment looking for a local signal would measure from these contributions is shown in fig.~\ref{fig:fNL} as a function of the redshift of the source $z_S$. As expected, the geometrical correction leads to an $\fNLloc \sim 1$ and roughly constant with $z_S$. Even though the redshift correction has the same $l$ dependence as the local signal, at low redshift it is enhanced  by a factor $(\etaS/\chiS)^2$ and the contamination to $\fNLloc$ becomes surprisingly large, $\fNLloc \gtrsim 10$ for $z_S \lesssim 0.3$. The overlap between the standard geometrical correction and the local signal strongly depends on the  redshift of the sources, and the $\fNLloc$ of contamination inherits the same behavior. In this case $\fNLloc \gtrsim 100$ for $z_S \gtrsim 0.8$. Finally, as the standard dynamical contribution gets larger and larger at low redshift, its contamination to the local signal does the same. For  $z_S \lesssim 1.7$ we have $\fNLloc \gtrsim 10^4$.

\clearpage
\section{Conclusion}
\label{sec:conclusion}

In this paper we present an exhaustive calculation of the full-sky bispectrum of the shear field as it would be built by an observer measuring shapes of galaxies at a given redshift. The calculation is performed at tree order (i.e.~neglecting loop corrections), assuming that the initial curvature perturbation $\zeta$ obeys Gaussian statistics, such as after single-field inflation. Furthermore, we study the relative importance of the different couplings and we compare them with primordial non-Gaussianities of the local type.

Our derivation fully exploits and completes the recent study of Ref.~\cite{Bernardeau:2009bm},  where the full-sky second-order expression of the shear field has been computed. This article derives all second-order terms in the metric fluctuations that contribute to the shear field and it is thus the natural starting point of the computation of the tree order bispectrum presented here. In this paper we focus our calculations on the bispectrum of the scalar, i.e.~electric, part of the shear field alone. This is somewhat restrictive since the second-order shear field contains a pseudo-scalar, i.e.~magnetic, mode. We leave the study of the $B$-mode and of the correlators where it is relevant for the future.

The computation of the bispectrum from the formal second-order shear expression is an involved exercise, which includes two steps. The first consists in the formal derivation of the bispectrum in harmonic space. The electric part of the shear field itself is defined from its harmonic space expression, in a way similar to the mode decompositions of the CMB polarization. Its bispectrum is then computed with the help of the decomposition properties of the spherical harmonics and their spin weighted extensions.  These results are presented in \sect~\ref{sec:BS} of the paper where the contribution of each term is fully computed. 
Then, the second part consists in numerically  evaluating the terms that we have found, in the context of the $\Lambda$CDM model.

Two approximations are commonly employed for weak lensing calculations, namely the Limber approximation \cite{1953ApJ...117..134L}, which consists in neglecting longitudinal modes, as they tend to average out along the line of sight, and the flat sky approximation, which consists in replacing a small portion of the spherical sky with a  2-dimensional plane.\footnote{Note that these two approximations are not equivalent and one does not necessarily imply the other.} In order to be consistent with our full-sky treatment, it is a priori not possible to employ neither of these two approximations.
Abandoning the Limber approximation makes the computations  numerically challenging due to the large number of integrations involved and, as a consequence, only a limited number of configurations could be exactly computed. 
In order to circumvent this problem we have used the higher-order Limber approximation discussed in \cite{1992ApJ...388..272K}. Apart from for a specific type of contributions, we have found that sufficiently high accuracy can be reached with the help of the second-order approximation.
The procedure that we have used to implement this approximation to the angular power spectrum and the bispectrum is presented in appendix~\ref{app:limber}. 

Our results are described in Sect. 5. In this section we give a precise account of the relative importance of the various contributing terms, as well as of their dependence with the multipole configuration. 
We distinguish the different contributing terms as geometrical and dynamical, depending on whether they originate from pure general relativistic effects on the line-of-sights or due to the dynamical evolution of the metric field. Note however that this distinction is somewhat arbitrary, since it depends on the gauge choice.
Furthermore, in each case we distinguish the standard terms as those which survive at small angular scale, from the non-standard ones that are mainly relevant at large angular scales.
 
We discuss the contribution to the bispectrum of the different nonlinear couplings focussing on the squeezed limit configuration, where the non-standard general relativistic corrections are more likely to play a significant role.
In \sect~\ref{sec:results} we give a precise account of the relative importance of the various contributing  terms as well as of their dependence with the multipole configuration.
We find that the general relativistic effects, i.e. the non-standard coupling terms,  are typically smaller than the standard non-linear couplings. However their relative importance increases at smaller source redshifts. Among the new couplings, the dynamical couplings are always much smaller than the others, while those due to the inhomogeneity in the redshift of the source dominate. The latter can even become 
of the same order of the standard couplings when the redshift of the sources is below $0.5$. These results are summarized in fig.~\ref{fig:bz1} and \ref{fig:bzp0}.

In \sect~\ref{sec:primordial} we finally compute the corresponding level of contamination induced by the standard and non-standard terms on the amplitude of primordial non-Gaussianities of the local type. 
Note that, as in the squeezed limit the standard non-linear couplings depend on the angle between the short and long mode, they differ significantly from a local signal. For $z_S \gtrsim0.8$, the contamination to $f_{\rm NL}^{\rm loc}$ of the standard geometrical couplings amounts to $f_{\rm NL}^{\rm loc} \gtrsim 100$. 
On the other hand, the relativistic corrections induce a non-Gaussianity which is mainly of the local type and can contaminate the search for a primordial signal by $f_{\rm NL}^{\rm loc} \sim 1$ to $f_{\rm NL}^{\rm loc}\gtrsim 10$ when the source redshifts vary from $1$ to $0.3$.

Our investigations did not go as far as computing the signal-to-noise ratio for the bispectrum induced by the general relativistic corrections. However, given the much larger amplitude of the standard contributions, we expect it to be rather small. 
This might not be the case when, besides the only $E$-mode three-point correlators we have investigated here, quantities involving $B$-modes are also included. Relativistic corrections indeed generically induce  $B$-modes with an amplitude a priori  comparable to the 
$E$-modes~\cite{Bernardeau:2009bm}. This is at variance with the standard dynamical couplings that induce second-order $E$-modes only. It ensures that the correlations between $E$ and $B$ modes are only non vanishing in the presence of general relativistic corrections making such quantities more likely to be observable. We leave such calculations for a future study.
%
\section*{Acknowledgements}
It is a pleasure to thank Chiara Caprini, Ruth Durrer, Ana\"\i s Rassat and Uros Seljak for useful discussions.
FB is partly supported by the French Programme National de Cosmologie et Galaxies; 
CB is supported by a Herchel Smith Postdoctoral Fellowship and by King's College Cambridge.

\vspace{1cm}

\hrule

\vspace{0.3cm}

\appendix

\section{Limber approximation}
 \label{app:limber}

\subsection{Power spectrum in second-order Limber approximation}
 \label{app:Cl_limber2}
The expression for the power spectrum with the second-order Limber approximation can be straightforwardly computed by applying the series expansion given by eq.~\eqref{Limber2_series} to the expression for the angular power spectrum, eq.~\eqref{C_l_3}. The result is
\[
\label{Cl_limber2}
C_l=
\nico^2(l,2)
\int_0^{\chiS} \frac{d\chi}{\chi}
T^2(\nu/\chi)P(\nu/\chi)
\tilde W(\chi)
\left[ \tilde W(\chi) -\frac{1}{\nu^2}
\left(\chi^2\tilde W''(\chi)+\frac13\chi^3\tilde W'''(\chi)
\right)\right]
\]
where $\tilde W(\chi)\equiv W(\chiS,\chi)/\sqrt\chi$. By dropping the $1/\nu^2$ term, one easily recovers the first-order Limber-approximated result given by eq.~\eqref{Cl_limber}.

\subsection{Geometrical terms in first-order Limber approximation}

We present here the solutions for the bispectrum at first order in Limber approximation. In our numerical code we go beyond this approximation and we use the second-order Limber approximation. However, the first-order approximation is sufficient to understand qualitatively the scaling of the bispectrum in the squeezed limit.

The bispectrum from the standard geometrical terms reads at first order in Limber approximation
\[
\label{bgeomstandlimb1}
\begin{split}
[b^{\rm (stan)}_{\rm geom}]_{l_1 l_2 l_3}   = \
&\nico(l_1,2)\nico(l_2,2)
\begin{pmatrix}l_1 & l_2 & l_3\\0 & 0 & 0\end{pmatrix}^{-1}
\\
&\Bigg\{\nico(l_1,3)\nico(l_2,1)
\left[
\begin{pmatrix}l_1 & l_2 & l_3\\3 &-1 & -2\end{pmatrix} +  c.c.\right]
\\
&\qquad
+\nico^2(l_2,1)\nico(l_1,2)
	\left[
	\begin{pmatrix}l_1 & l_2 & l_3\\2 &0 & -2\end{pmatrix} +  c.c. \right]
\\
&\qquad
+\nico^2(l_1,1)\nico(l_2,2)
	\left[
	\begin{pmatrix}l_1 & l_2 & l_3\\0 &2 & -2\end{pmatrix} +  c.c. \right]
\\
&\qquad
+\nico(l_1,1)\nico(l_2,1)(l_1+2)(l_1-1)
	\left[\begin{pmatrix}l_1 & l_2 & l_3\\1 &1 & -2\end{pmatrix} +  c.c. \right]
	\Bigg\}	
\\
&\times  \Bigg[
\int_0^{\chiS}\frac{d\chi}{\chi^3}
T^2\!\!\left(\frac{\nu_1}{\chi}\right)P\!\left(\frac{\nu_1}{\chi}\right)
W^2(\chiS,\chi)
\int_0^{\chi}\frac{d\chi'}{\chi'^2}
T^2\!\!\left(\frac{\nu_2}{\chi'}\right)P\!\left(\frac{\nu_2}{\chi'}\right)
W(\chiS,\chi')g(\chi')
\\
&\qquad
-\int_0^{\chiS}\frac{d\chi}{\chi^2}
T^2\!\!\left(\frac{\nu_1}{\chi}\right)P\!\left(\frac{\nu_1}{\chi}\right)
W^2(\chiS,\chi)
\int_0^{\chi}\frac{d\chi'}{\chi'^3}
T^2\!\!\left(\frac{\nu_2}{\chi'}\right)P\!\left(\frac{\nu_2}{\chi'}\right)
W^2(\chiS,\chi') g(\chi')
\Bigg]
\\
+
& \   \nico^2(l_1,1) \nico(l_1,2) \nico^2(l_2,2)
\left[
\begin{pmatrix}l_1 & l_2 & l_3\\0 &2 & -2\end{pmatrix} +  c.c.
\right]
\begin{pmatrix}l_1 & l_2 & l_3\\0 & 0 & 0\end{pmatrix}^{-1}
\\
& \times
\int_0^{\chiS}\frac{d\chi}{\chi^2}
T^2\!\!\left(\frac{\nu_1}{\chi}\right)P\!\left(\frac{\nu_1}{\chi}\right)
W^2(\chiS,\chi)
\int_0^{\chiS}\frac{d\chi'}{\chi'^2}
T^2\!\!\left(\frac{\nu_2}{\chi'}\right)P\!\left(\frac{\nu_2}{\chi'}\right)
W^2(\chiS,\chi')
\\
+&  \ 5 \ {\rm perms}\;,
\end{split}
\]
where $\nu_i\equiv l_i+1/2$.

The bispectrum from the geometrical corrections is at first order in Limber approximation
\[
\label{bgeomcorrlimb1}
\begin{split}
[b^{\rm (corr)}_{\rm geom}]_{l_1 l_2 l_3}   = \
& \nico(l_1,2)\nico(l_2,2)\nico(l_3,2)
\\
&
\Bigg\{\int_0^{\chiS}\!\!\!\frac{d\chi}{\chi^4}
T^2\!\!\left(\frac{\nu_1}{\chi}\right)P\!\left(\frac{\nu_1}{\chi}\right)
T^2\!\!\left(\frac{\nu_2}{\chi}\right)P\!\left(\frac{\nu_2}{\chi}\right)
W^3(\chiS,\chi)g(\chi)
\\
&\quad+ 2\int_0^{\chiS}\!\!\!\frac{d\chi}{\chi^3}
T^2\!\!\left(\frac{\nu_1}{\chi}\right)P\!\left(\frac{\nu_1}{\chi}\right)
W(\chiS,\chi)
\\
&\qquad\times \Bigg[ -\frac{g(\chi)}{\chiS}
\int_0^{\chi}\frac{d\chi'}{\chi'^2}
T^2\!\!\left(\frac{\nu_2}{\chi'}\right)P\!\left(\frac{\nu_2}{\chi'}\right)
W(\chiS,\chi')g(\chi')
\\
&\qquad\qquad+g'(\chi)
\int_0^\chi \frac{d\chi'}{\chi'^2}
T^2\!\!\left(\frac{\nu_2}{\chi'}\right)P\!\left(\frac{\nu_2}{\chi'}\right)
W(\chiS,\chi')g(\chi')
\Bigg]
\Bigg\}
\\
- & \ 2 \nico(l_1,2) \nico^2(l_2,2)
\left[
\begin{pmatrix}l_1 & l_2 & l_3\\0 & 2 & -2\end{pmatrix} +  c.c.
\right]
\begin{pmatrix}l_1 & l_2 & l_3\\0 & 0 & 0\end{pmatrix}^{-1}
\\
&\quad\times\int_0^{\chiS}\frac{d\chi}{\chi^3}
T^2\!\!\left(\frac{\nu_1}{\chi}\right)P\!\left(\frac{\nu_1}{\chi}\right)
W(\chiS,\chi)
\\
& \quad\times \Bigg[ W(\chiS,\chi)
\int_0^\chi \frac{d\chi'}{\chi'^2}
T^2\!\!\left(\frac{\nu_2}{\chi'}\right)P\!\left(\frac{\nu_2}{\chi'}\right)
W(\chiS,\chi')g(\chi')
\\
&\qquad\qquad+\frac{\chi}{\chiS}g(\chi)
\int_0^\chi \frac{d\chi'}{\chi'^3}
T^2\!\!\left(\frac{\nu_2}{\chi'}\right)P\!\left(\frac{\nu_2}{\chi'}\right)
W(\chiS,\chi')g(\chi')
\\
&\qquad\qquad - \frac{g(\chi)\chi}{\chiS}
\int_0^{\chiS}\frac{d\chi'}{\chi'^2}
T^2\!\!\left(\frac{\nu_2}{\chi'}\right)P\!\left(\frac{\nu_2}{\chi'}\right)
W^2(\chiS,\chi')\Bigg]
\\
+& \ 5 \ {\rm perms}\;.
\end{split}
\]

The bispectrum from the redshift corrections is at first order in Limber approximation
\[
\label{bzcorrlimb1}
\begin{split}
[b^{\rm (corr)}_{\rm z}]_{l_1 l_2 l_3} = \ &
\frac12\nico^2(l_1,2)\nico(l_2,2)
\left[
\begin{pmatrix}l_1 & l_2 & l_3\\2 & 0 & -2\end{pmatrix} + c.c. \right]
\begin{pmatrix}l_1 & l_2 & l_3\\0 & 0 & 0\end{pmatrix}^{-1}
\\
& \times\int_0^{\chiS}\frac{d\chi}{\chi^2}\frac{d\chi'}{\chi'^2}
P\!\left(\frac{\nu_1}{\chi}\right)T^2\!\left(\frac{\nu_1}{\chi}\right)
P\!\left(\frac{\nu_2}{\chi'}\right)T^2\!\left(\frac{\nu_2}{\chi'}\right)
W(\chiS,\chi)W(\chiS,\chi')
\\
& \times\Bigg\{
\frac{1}{\chiS^2\Hconf_S}
\left[2g'(\chi')
-\nu_2^{3/2}\sqrt{\frac2\pi}\ g(\chiS)
\ \frac{1}{\chi'}\ j_{l_2}\!\left(\frac{\nu_2}{\chi'}\chiS\right)
\right]
\\
&\qquad +\frac{2 a_S }{3H_0^2\Omega_m\chiS^2}
\left[g(\chiS)-\frac{g'(\chiS)}{\Hconf_S}\right]\nu_2^{5/2}
\sqrt{\frac2\pi}\ \frac{1}{\chi'^2}\
j_{l_2}'\!\left(\frac{\nu_2}{\chi'}\chiS\right)\Bigg\}
\\
+& \ 5 \ {\rm perms}\;.
\end{split}
\]
Note that for the redshift corrections we cannot use the Limber approximation
on the spherical Bessel functions $j_{l_2}\left(\frac{\nu_2}{\chi'}\chi_S\right)$ and
$j'_{l_2}\left(\frac{\nu_2}{\chi'}\chi_S\right)$. Indeed in this approximation
$\chi'\simeq \chi_S$ and the bispectrum vanishes. Hence for these terms we
perform the exact integral on $\chi'$.

\subsection{Standard dynamical term in first-order Limber approximation}

By applying the first-order Limber approximation given by eq.~\eqref{limber1_dubble_jl} to eq.~\eqref{bdynnewt}, it is straightforward to compute the Limber-approximated expression of the standard dynamical term.
\[
\begin{split}
[b^{\rm (stan)}_{\rm dyn}]_{l_1 l_2 l_3}  =&\
\nico(l_1,2) \nico(l_2,2) \nico(l_3,2)
\frac{\nu_1^2\nu_2^2}{\nu_3^2}
\frac{2}{3\Omega_m H_0^2}
\\
&
\int_0^{\chiS} \frac{d \chi}{\chi^6} W^3(\chiS,\chi)g(\chi)a(\chi)
F_{2,N}\left(\frac{\nu_1}{\chi},\frac{\nu_2}{\chi},\frac{\nu_3}{\chi};\chi\right)
T^2\left(\frac{\nu_1}{\chi}\right) P\left(\frac{\nu_1}{\chi}\right)
T^2\left(\frac{\nu_2}{\chi}\right) P\left(\frac{\nu_2}{\chi}\right)
\\
+& \ 5 \ {\rm perms}\;.
\end{split}
\]

\section{Dynamical contributions}
 \label{app:dyn}

We present here the computation of the dynamical scalar and tensor terms. The vector terms are shown in section~\ref{sec:dynamical}.

Let us start by the scalar terms. The sum of the two relativistic potentials gives, using eqs.~\eqref{phi2} and \eqref{psi2},
\[
\frac12 \left(\phir(\vec k_3)+\psir(\vec k_3)\right) ={T(k_1) T(k_2)} F_{2,S} (\vec k_1, \vec k_2,\vec k_3) \Phi_{\vec k_1} \Phi_{\vec k_2} \;,
\]
where we define
\[
 F_{2,S} (k_1,k_2,k_3) \equiv
 \frac{ \vec k_1 \cdot \vec k_2 - 3 {(\hat k_3 \cdot \vec k_1)(\hat k_3 \cdot \vec k_2)} }{ 6 k_3^2} \;.
\]
For the bispectrum one finds
\[
\label{bdyncorrscal}
\begin{split}
[b^{{\rm (corr)}}_{{\rm dyn}, S}]_{l_1 l_2 l_3}  =
& - \nico(l_1,2) \nico(l_2,2) \nico(l_3,2)
\int_0^{\chiS}  d\chi_1 d\chi_2 d\chi_3
W(\chiS,\chi_1)W(\chiS,\chi_2)  W(\chiS,\chi_3)
\\
& \times
\int \frac{2 k_1^2dk_1}{\pi} \frac{2 k_2^2dk_2}{\pi} \frac{2 k_3^2dk_3}{\pi}
T^2(k_1)T^2(k_2) F_{2,S}(k_1,k_2,k_3) P(k_1) P(k_2)
\\
&\times
\int_0^{\infty} d \chi \chi^2
j_{l_1}( k_1 \chi_1) j_{l_1}( k_1 \chi)
j_{l_2}( k_2 \chi_2) j_{l_2}( k_2 \chi)
j_{l_3}( k_3 \chi_3) j_{l_3}( k_3 \chi)
\\
+&\ \hbox{5 perms} \; .
\end{split}
\]

We now compute the tensor terms. We start with the contribution coming from the first line of eq.~\eqref{gamma_corr_dyn}. 
Using eq.~\eqref{gamma}, after few manipulations one finds
\[
\begin{split}
- \frac14 h_{rr} (k_3)=
& - \frac{5}{3}\left[ 1 - 3 \frac{j_1(k_3\eta)}{k_3 \eta} \right]
\frac{{T(k_1) T(k_2)}}{k_3^4}  \Phi_{\vec k_1} \Phi_{\vec k_2}  \Bigg[\frac{1}{8}\Big(k_3^4+(k_1^2-k_2^2)^2-2k_3^2(k_1^2+k_2^2)\Big)
\left(1+(\hat k_3\cdot \hat n)^2\right)
\\
&+(\vec k_3\cdot \hat n)\Big( k_2^2(\vec k_1\cdot \hat n)+k_1^2(\vec k_2\cdot \hat n) \Big)
- k_3^2 (\vec k_1\cdot \hat n)(\vec k_2\cdot \hat n) \Bigg]  \;.
\end{split}
\]
The computation of this contribution is similar to the one of the vector contribution, eq.~\eqref{vecrr}. The only new term is the one proportional to the product $(\vec k_1 \cdot \hat n)(\vec k_2 \cdot \hat n)$. We use the expansion of eq.~(\ref{exp_qi}) twice. As a result, in the computation of the $a_{E,lm}$ we obtain an integral over $d^3 \vec k_3$ with four spherical harmonics.
This integral gives rise to a product of two Gaunt integrals. We then compute the bispectrum regrouping the appropriate Gaunt integrals such that the summation in eq.~(\ref{sumgaunt}) can be applied twice. This generates two Wigner 6-$j$ symbols, with terms of the form $l_i\pm 2$. With this the bispectrum from the tensor contribution $- \frac14 h_{rr}$ is

\[
\begin{split}
\label{bdyncorrtenslong}
[b_{\rm dyn}^{(\rm corr)}]_{l_1 l_2 l_3}=
&\frac{5}{3}\nico(l_1,2)\nico(l_2,2)\nico(l_3,2)
\int_0^{\chiS}d\chi_1 d\chi_2 d\chi_3 W(\chi_1,\chiS)W(\chi_2,\chiS)W(\chi_3,\chiS)
\\
& \times \int \frac{2 k_1^2dk_1}{\pi} \frac{2 k_2^2dk_2}{\pi} \frac{2 k_3^2dk_3}{\pi} T^2(k_1)T^2(k_2)
P(k_1) P(k_2)\left[1-\frac{3j_1\big(k_3(\eta_0-\chi_3)\big)}{k_3(\eta_0-\chi_3)}\right]
\\
&\times\Bigg\{ \frac{k_3^4+(k_1^2-k_2^2)^2-2k_3^2(k_1^2+k_2^2)}{8k_3^4}
\\
&\qquad\qquad\times
\int_0^{\infty} d \chi \chi^2
j_{l_1}(k_1\chi_1)j_{l_1}( k_1 \chi)j_{l_2}(k_2\chi_2)j_{l_2}( k_2 \chi)
\left(1-\frac{\partial^2_{\chi_3}}{k_3^2}\right)j_{l_3}(k_3\chi_3)j_{l_3}( k_3 \chi)\\
&\qquad+2\frac{k_1k_2^2}{k_3^4}
\sum_{\substack{l'_1=l_1\pm 1\\l'_3=l_3\pm1}}(-1)^{l_2}i^{l_1'-l_1+1}(2l_1'+1)(2l_3'+1)
\\
&\qquad\qquad\times
\begin{Bmatrix}l_1&l_2&l_3\\l'_3 &1&l'_1\end{Bmatrix}
\begin{pmatrix}l_1 & l'_1 & 1\\0 & 0 & 0\end{pmatrix}
\begin{pmatrix}l_3 & l'_3 & 1\\0 & 0 &0\end{pmatrix}
\begin{pmatrix}l'_1 & l_2 & l'_3 \\ 0 & 0 & 0\end{pmatrix}
\begin{pmatrix}l_1 & l_2 & l_3 \\ 0 & 0 & 0\end{pmatrix}^{-1}
\\
&\qquad\qquad\times
\int_0^{\infty} d \chi \chi^2
j_{l_1}(k_1\chi_1)j_{l'_1}( k_1 \chi)j_{l_2}(k_2\chi_2)j_{l_2}( k_2 \chi)
\partial_{\chi_3}j_{l'_3}(k_3\chi_3)j_{l'_3}( k_3 \chi)
\\
&\qquad+\frac{k_1k_2}{k_3^2}
\sum_{\substack{l'_1=l_1\pm 1\\l_2'=l_2\pm 1}}
\sum_{\substack{l'_3=l_3\pm1\\l''_3=l'_3\pm 1}}
(-1)^{l_3}i^{l_1'-l_1+l'_2-l_2}(2l_1'+1)(2l_2'+1)(2l_3'+1)(2l_3''+1)
\\
&
\qquad\qquad\times
\begin{Bmatrix}l_1&l_3&l_2\\1 &l'_2&l'_3\end{Bmatrix}
\begin{Bmatrix}l_1&l'_2&l'_3\\l''_3 &1&l'_1\end{Bmatrix}
\begin{pmatrix}l_1 & l'_1 & 1\\0 & 0 & 0\end{pmatrix}
\begin{pmatrix}l_2 & l'_2 & 1\\0 & 0 & 0\end{pmatrix}
\\
&\qquad\qquad\times
\begin{pmatrix}l_3 & l'_3 & 1\\0 & 0 &0\end{pmatrix}
\begin{pmatrix}l'_1 & l'_2 & l''_3 \\ 0 & 0 & 0\end{pmatrix}
\begin{pmatrix}l_1 & l_2 & l_3 \\ 0 & 0 & 0\end{pmatrix}^{-1}
\\
&\qquad\qquad\times
\int_0^{\infty} d \chi \chi^2
j_{l_1}(k_1\chi_1)j_{l'_1}( k_1 \chi)j_{l_2}(k_2\chi_2)j_{l'_2}( k_2 \chi)
j_{l''_3}(k_3\chi_3)j_{l''_3}( k_3 \chi)\Bigg\}
\\
&+\ \hbox{5 perms}\; .
\end{split}
\]

We then compute the tensor term in the second line of eq.~\eqref{gamma_corr_dyn} that reads
\[
\begin{split}
- \frac12 \spart {}_1 h_r(\hat n \chi)  =
& - \frac{10}{3}
\int \frac{d^3\vec k_3}{(2\pi)^3}\frac{d^3\vec k_1 d^3\vec k_2}{(2\pi)^3}
\delta(\vec k_3-\vec k_1-\vec k_2)
\left[ 1 - 3 {j_1(k_3\eta)}/({k_3 \eta}) \right]
\frac{{T(k_1) T(k_2)}}{k_3^4}  \Phi_{\vec k_1} \Phi_{\vec k_2}
\\
&\times \spart \Bigg\{e^{i\vec k_3\cdot \hat n \chi}
\Bigg[\frac{1}{8}
\Big(k_3^4+(k_1^2-k_2^2)^2-2k_3^2(k_1^2+k_2^2)\Big)
(\hat k_3\cdot \hat n)(\hat k_3\cdot \hat e_+)
\\
&\qquad\qquad
+k_1k_2k_3(\hat k_3\cdot \hat e_+)
\Big( k_2(\hat k_1\cdot \hat n)+k_1(\hat k_2\cdot \hat n) \Big)
\\
&\qquad\qquad
+\frac12 k_1k_2(k_2^2-k_1^2-k_3^2)
\Big(
(\hat k_1\cdot \hat e_+)(\hat k_2\cdot \hat n)
-(\hat k_2\cdot \hat e_+)(\hat k_1\cdot \hat n)
\Big)\Bigg]\Bigg\}  \;.
\end{split}
\]

The computation of this term is similar to the one of the vector term in eq.~\eqref{vectr} and we find for the bispectrum

\[
\begin{split}
\label{bdyncorrtenstr}
[b_{\rm dyn}^{(\rm corr)}]_{l_1 l_2 l_3}=
&\frac{10}{3}\nico(l_1,2)\nico(l_2,2)\nico(l_3,2)
\int_0^{\chiS}d\chi_1 d\chi_2 d\chi_3
W(\chi_1,\chiS)W(\chi_2,\chiS)\frac{1}{\chi_3^2}
\\
& \times \int \frac{2 k_1^2dk_1}{\pi} \frac{2 k_2^2dk_2}{\pi} \frac{2 k_3^2dk_3}{\pi}
T^2(k_1)T^2(k_2) P(k_1) P(k_2)
\left[1-\frac{3j_1\big(k_3(\eta_0-\chi_3)\big)}{k_3(\eta_0-\chi_3)}\right]
\\
&\times\Bigg\{ \frac{k_3^4+(k_1^2-k_2^2)^2-2k_3^2(k_1^2+k_2^2)}{8k_3^5}
\sum_{l'_3=l_3\pm 1}i^{l'_3-l_3+1}\frac{2l'_3+1}{2l_3+1}
\begin{pmatrix}l_3 & l'_3 & 1\\0 & 0 &0\end{pmatrix}
\begin{pmatrix}l_1 & l_2 & l_3 \\ 0 & 0 & 0\end{pmatrix}^{-1}
\\
&\qquad \times \Bigg[(l_3^{'2}+l'_3+2)(l'_3+1)l'_3
\begin{pmatrix}l_3 & l'_3 & 1\\0 & 0 &0\end{pmatrix}
+\sqrt{2l'_3(l'_3+1)}(3l_3^{'2}+3l'_3-2)
\begin{pmatrix}l_3 & l'_3 & 1\\0 & 1 &-1\end{pmatrix} \Bigg]
\\
&\qquad \times
\int_0^{\infty} d \chi \chi^2
j_{l_1}(k_1\chi_1)j_{l_1}( k_1 \chi)j_{l_2}(k_2\chi_2)j_{l_2}( k_2 \chi)
j_{l_3}(k_3\chi_3)j_{l'_3}( k_3 \chi)
\\
&\quad-\frac{k_1k_2^2}{k_3^4}
\sum_{\substack{l'_1=l_1\pm 1\\l'_3=l_3\pm1}}(-1)^{l_2}i^{l_1'-l_1+1}
\frac{(2l_1'+1)(2l_3'+1)}{\nico(l_3,2)^2}
\\
&
\qquad\times
\begin{Bmatrix}l_1&l_3&l_2\\l'_3 &l'_1&1\end{Bmatrix}
\begin{pmatrix}l_1 & l'_1 & 1\\0 & 0 &0\end{pmatrix}
\begin{pmatrix}l'_1 & l_2 & l'_3\\0 & 0 &0\end{pmatrix}
\begin{pmatrix}l_1 & l_2 & l_3 \\ 0 & 0 & 0\end{pmatrix}^{-1}
\\
&\qquad \times \Bigg[(l_3^{'2}+l'_3+2)(l'_3+1)l'_3
\begin{pmatrix}l_3 & l'_3 & 1\\0 & 0 &0\end{pmatrix}
+\sqrt{2l'_3(l'_3+1)}(3l_3^{'2}+3l'_3-2)
\begin{pmatrix}l_3 & l'_3 & 1\\0 & 1 &-1\end{pmatrix}\Bigg]
\\
&\qquad\times \int_0^{\infty} d \chi \chi^2j_{l_1}(k_1\chi_1)j_{l'_1}( k_1 \chi)j_{l_2}(k_2\chi_2)j_{l_2}( k_2 \chi)
j_{l'_3}(k_3\chi_3)j_{l'_3}( k_3 \chi)
\\
&\quad-\chi_3\frac{(k_3^2+k_1^2-k_2^2)k_1k_2}{k_3^4}\sum_{\substack{l'_1=l_1\pm 1\\l_2'=l_2\pm 1}}\sum_{\substack{l'_3=l_3\pm1\\l''_3=l'_3\pm 1}}
(-1)^{l_3}\frac{i^{l_1'-l_1+l'_2-l_2}}{\nico(l_3,2)^2}(2l_1'+1)(2l_2'+1)(2l_3'+1)(2l_3''+1)
\\
&\qquad\times
\begin{Bmatrix}l_1&l_3&l_2\\1 &l'_2&l'_3\end{Bmatrix}
\begin{Bmatrix}l_1&l'_2&l'_3\\l''_3 &1&l'_1\end{Bmatrix}
\begin{pmatrix}l_1 & l'_1 & 1\\0 & 0 & 0\end{pmatrix}
\begin{pmatrix}l_2 & l'_2 & 1\\0 & 0 & 0\end{pmatrix}
\begin{pmatrix}l'_1 & l'_2 & l''_3 \\ 0 & 0 & 0\end{pmatrix}
\begin{pmatrix}l_1 & l_2 & l_3 \\ 0 & 0 & 0\end{pmatrix}^{-1}
\\
&\qquad\times \Bigg[4(l_3^{''2}+l''_3+2)
\begin{pmatrix}l_3 & l'_3 & 1\\0 & 0 &0\end{pmatrix}
\begin{pmatrix}l'_3 & l''_3 & 1\\0 & 0 &0\end{pmatrix}
\\
&\qquad\qquad+4(l_3^{''2}+l''_3+2)
\begin{pmatrix}l_3 & l'_3 & 1\\0 & 1 &-1\end{pmatrix}
\begin{pmatrix}l'_3 & l''_3 & 1\\-1 & 0 &1\end{pmatrix}
\\
&\qquad\qquad+\sqrt{2l''_3(l''_3+1)}(l_3^{''2}+l''_3+6)
\begin{pmatrix}l_3 & l'_3 & 1\\0 & 0 &0\end{pmatrix}
\begin{pmatrix}l'_3 & l''_3 & 1\\0 & -1 &1\end{pmatrix}
\\
&\qquad\qquad+8\sqrt{2l''_3(l''_3+1)}
\begin{pmatrix}l_3 & l'_3 & 1\\0 & 1 &-1\end{pmatrix}
\begin{pmatrix}l'_3 & l''_3 & 1\\-1 & 1 &0\end{pmatrix}
\\
&\qquad\qquad+2\nico(l''_3,2)
\begin{pmatrix}l_3 & l'_3 & 1\\0 & -1 &1\end{pmatrix}
\begin{pmatrix}l'_3 & l''_3 & 1\\1 & -2 &1\end{pmatrix}\Bigg]
\\
&\qquad\times \int_0^{\infty} d \chi \chi^2j_{l_1}(k_1\chi_1)j_{l'_1}( k_1 \chi)j_{l_2}(k_2\chi_2)j_{l'_2}( k_2 \chi)
j_{l''_3}(k_3\chi_3)j_{l''_3}( k_3 \chi)\Bigg\}
\\
+&\ \hbox{5 perms}\; .
\end{split}
\]

Finally we compute the bispectrum of the boundary tensor term, i.e. the last term in eq.~\eqref{gamma_corr_dyn}
\[
\begin{split}
- \frac14 {}_2 h (\hat n \chiS)  =
&- \frac{5}{6}\int \frac{d^3\vec k_3}{(2\pi)^3}\frac{d^3\vec k_1d^3\vec k_2}{(2\pi)^3}
\delta(\vec k_3-\vec k_1-\vec k_2)
\left[ 1 - 3 {j_1(k_3\eta)}/({k_3 \eta}) \right]
\frac{{T(k_1) T(k_2)}}{k_3^4}
\Phi_{\vec k_1} \Phi_{\vec k_2}
\\
&\times e^{i\vec k_3\cdot \hat n \chiS}
\Bigg[\frac{1}{4}
\Big(k_3^4+(k_1^2-k_2^2)^2-2k_3^2(k_1^2+k_2^2)\Big)
(\hat k_3\cdot \hat e_+)^2
\\
&\phantom{\times e^{i\vec k_3\cdot \hat n \chiS}\Bigg[}
+2(\vec k_3\cdot \hat e_+)
\Big( k_2^2(\vec k_1\cdot \hat e_+)+k_1^2(\vec k_2\cdot \hat e_+) \Big)
-2 k_3^2(\vec k_1\cdot \hat e_+)
(\vec k_2\cdot \hat e_+)\Bigg]  \;.
\end{split}
\]
The computation of this term is similar to the previous ones and its bispectrum reads
\[
\begin{split}
\label{bdyncorrtenssource}
[b_{\rm dyn}^{(\rm corr)}]_{l_1 l_2 l_3}=
&\frac{5}{3}\nico(l_1,2)\nico(l_2,2)\nico(l_3,2)
\int_0^{\chiS\ }d\chi_1\, d\chi_2\,  W(\chi_1,\chiS)\, W(\chi_2,\chiS)
\\
\times& \int \frac{2 k_1^2dk_1}{\pi} \frac{2 k_2^2dk_2}{\pi} \frac{2 k_3^2dk_3}{\pi} T^2(k_1)T^2(k_2)
P(k_1) P(k_2)\left[1-\frac{3j_1\big(k_3(\eta_0-\chiS)\big)}{k_3(\eta_0-\chiS)}\right]
\\
\times&\Bigg\{-\frac{k_3^4+(k_1^2-k_2^2)^2-2k_3^2(k_1^2+k_2^2)}{16k_3^4}
\frac{1}{k_3^2\chiS^2}
\\
& \qquad\times\int_0^{\infty} d \chi \chi^2j_{l_1}(k_1\chi_1)j_{l_1}( k_1 \chi)j_{l_2}(k_2\chi_2)j_{l_2}( k_2 \chi)j_{l_3}(k_3\chiS)j_{l_3}( k_3 \chi)
\\
&\phantom{\Bigg\{}+\frac{2k_1k_2^2}{k_3^3}\sum_{\substack{l'_1=l_1\pm 1\\l'_3=l_3\pm1}}(-1)^{l_2}i^{l_1'-l_1+1}\frac{(2l_1'+1)(2l_3'+1)}{\nico(l_3,2)^2}
\\
&\qquad\times
\begin{Bmatrix}l_1&l_3&l_2\\l'_3 &l'_1&1\end{Bmatrix}
\begin{pmatrix}l_1 & l'_1 & 1\\0 & 0 &0\end{pmatrix}
\begin{pmatrix}l'_1 & l_2 & l'_3\\0 & 0 &0\end{pmatrix}
\begin{pmatrix}l_1 & l_2 & l_3 \\ 0 & 0 & 0\end{pmatrix}^{-1}
\\
&\qquad\times \Bigg[4(l'_3+1)l'_3
\begin{pmatrix}l_3 & l'_3 & 1\\0 & 0 &0\end{pmatrix}
+\sqrt{2l'_3(l'_3+1)}(l_3'^{2}+l'_3+2)
\begin{pmatrix}l_3 & l'_3 & 1\\0 & 1 &-1\end{pmatrix}\Bigg]
\\
&\qquad\times\frac{1}{k_3\chiS} \int_0^{\infty} d \chi \chi^2j_{l_1}(k_1\chi_1)j_{l'_1}( k_1 \chi)j_{l_2}(k_2\chi_2)j_{l_2}( k_2 \chi)
j_{l'_3}(k_3\chi_3)j_{l'_3}( k_3 \chi)
\\
&\phantom{\Bigg\{}+\frac{2k_1k_2}{k_3^2}
\sum_{\substack{l'_1=l_1\pm 1\\l_2'=l_2\pm 1}}\sum_{\substack{l'_3=l_3\pm1\\l''_3=l'_3\pm 1}}
(-1)^{l_3}\frac{i^{l_1'-l_1+l'_2-l_2}}{\nico(l_3,2)^2}(2l_1'+1)(2l_2'+1)(2l_3'+1)(2l_3''+1)
\\
&\qquad\times
\begin{Bmatrix}l_1&l_3&l_2\\1 &l'_2&l'_3\end{Bmatrix}
\begin{Bmatrix}l_1&l'_2&l'_3\\l''_3 &1&l'_1\end{Bmatrix}
\begin{pmatrix}l_1 & l'_1 & 1\\0 & 0 & 0\end{pmatrix}
\begin{pmatrix}l_2 & l'_2 & 1\\0 & 0 & 0\end{pmatrix}
\begin{pmatrix}l'_1 & l'_2 & l''_3 \\ 0 & 0 & 0\end{pmatrix}
\begin{pmatrix}l_1 & l_2 & l_3 \\ 0 & 0 & 0\end{pmatrix}^{-1}
\\
&\qquad \times \Bigg[4
\begin{pmatrix}l_3 & l'_3 & 1\\0 & 1 &-1\end{pmatrix}
\begin{pmatrix}l'_3 & l''_3 & 1\\-1 & 0 &1\end{pmatrix}
+2\sqrt{2l''_3(l''_3+1)}
\begin{pmatrix}l_3 & l'_3 & 1\\0 & 0 &0\end{pmatrix}
\begin{pmatrix}l'_3 & l''_3 & 1\\0 & -1 &1\end{pmatrix}
\\
&\qquad\qquad+2\sqrt{2l''_3(l''_3+1)}
\begin{pmatrix}l_3 & l'_3 & 1\\0 & 1 &-1\end{pmatrix}
\begin{pmatrix}l'_3 & l''_3 & 1\\-1 & 1 &0\end{pmatrix}
\\
&\qquad\qquad
+\nico(l''_3,2)
\begin{pmatrix}l_3 & l'_3 & 1\\0 & -1 &1\end{pmatrix}
\begin{pmatrix}l'_3 & l''_3 & 1\\1 & -2 &1\end{pmatrix}\Bigg]
\\
& \qquad\times\int_0^{\infty} d \chi \chi^2
j_{l_1}(k_1\chi_1)j_{l'_1}( k_1 \chi)j_{l_2}(k_2\chi_2)j_{l'_2}( k_2 \chi)
j_{l''_3}(k_3\chiS)j_{l''_3}( k_3 \chi)\Bigg\}
\\
+&\ \hbox{5 perms} \; .
\end{split}
\]

\section{Spin-weighted spherical harmonics}
 \label{app:spart}

The spin-weighted spherical harmonics ${}_s Y_{lm}(\hat n)$ are related to the standard spherical harmonics $Y_{lm}(\hat n)$ by
\[
\nico(l,s){}_s Y_{lm}(\hat n) = \dslash^s Y_{lm}(\hat n) \;,\label{def_sw_1}
\]
\[
\nico(l,s){}_{-s} Y_{lm}(\hat n) = (-1)^s \dslashbar^s Y_{lm}(\hat n) \;, \label{def_sw_2}
\]
for $s\geq0$ and $|s|\leq l$. The coefficient $\nico(l,s)$ is defined as
\[
\nico(l,s)\equiv\sqrt{\frac{(l+s)!}{(l-s)!}} \;.
\]

Let us recall here a few useful relations obeyed by the spin-weighted spherical harmonics,
\begin{align}
{}_s Y_{lm}^*(\hat n) &= (-1)^{m+s} {}_{-s} Y_{l,-m}(\hat n) \;, \\
\dslashbar\dslash Y_{lm}(\hat n) &=-\nico^2(l,1) Y_{lm} = -l(l+1) Y_{lm}(\hat n)\;,\\
\int d\hat n\ {}_s Y_{lm}(\hat n)\ {}_s Y_{l'm'}^*(\hat n) &= \delta_{ll'}\delta_{mm'}\;,\\
\sum_{lm} {}_s Y_{lm}(\hat n) {}_s Y_{lm}^*(\hat n') &= \delta(\hat n-\hat n')\;.
\end{align}

The operators $\dslash$ and $\dslashbar$ satisfy the commutation relation
\[
(\dslashbar\dslash-\dslash\dslashbar)\ {}_s Y_{lm} = 2s\ {}_s Y_{lm} \;,
\]
such that
\[
\dslashbar\dslash^2\ Y_{lm}=-\frac{\nico^2(l,2)}{\nico^2(l,1)}\dslash\ Y_{lm} \;.
\]
The same relation holds when $\dslash$ and $\dslashbar$ are interchanged. From this relation we deduce that
\[
\dslashbar^2\dslash^2\ Y_{lm} = \nico^2(l,2)\ Y_{lm} \;.
\]
Dropping the boundary term when using integration by parts is valid as long as the integrand has spin zero, i.e. $\int d\hat n\ \dslash([\mathrm{spin=-1}])=\int d\hat n\ \dslashbar([\mathrm{spin=1}])=0$.


\bibliographystyle{plain}
\bibliography{Lens13}

\end{document}